\documentclass[iop]{emulateapj}

\def\lapp{\ifmmode\stackrel{<}{_{\sim}}\else$\stackrel{<}{_{\sim}}$\fi}
\def\gapp{\ifmmode\stackrel{>}{_{\sim}}\else$\stackrel{>}{_{\sim}}$\fi}
\usepackage{multirow}
\usepackage{color}

\newcommand{\AR}{\textcolor{black}}

\begin{document}

\title{High-Energy X-ray Imaging of the Pulsar Wind Nebula  MSH~15$-$5{\sl2}: Constraints on Particle Acceleration and Transport}

\author{
Hongjun An\altaffilmark{1},
Kristin K. Madsen\altaffilmark{2},
Stephen P. Reynolds\altaffilmark{3},
Victoria M. Kaspi\altaffilmark{1},
Fiona A. Harrison\altaffilmark{2},
Steven~E.~Boggs\altaffilmark{4},
Finn E. Christensen\altaffilmark{5}, William~W.~Craig\altaffilmark{4,6},
Chris L. Fryer\altaffilmark{7},
Brian W. Grefenstette\altaffilmark{2}, Charles J. Hailey\altaffilmark{8},
Kaya Mori\altaffilmark{8},
Daniel Stern\altaffilmark{9}, and William W. Zhang\altaffilmark{10}\\
}

\affil{
{\small $^1$Department of Physics, McGill University, Montreal, Quebec, H3A 2T8, Canada}\\
{\small $^2$Cahill Center for Astronomy and Astrophysics, California Institute of Technology, Pasadena, CA 91125, USA}\\
{\small $^3$Physics Department, NC State University, Raleigh, NC 27695, USA}\\
{\small $^4$Space Sciences Laboratory, University of California, Berkeley, CA 94720, USA}\\
{\small $^5$DTU Space, National Space Institute, Technical University of Denmark, Elektrovej 327, DK-2800 Lyngby, Denmark}\\
{\small $^6$Lawrence Livermore National Laboratory, Livermore, CA 94550, USA}\\
{\small $^7$CCS-2, Los Alamos National Laboratory, Los Alamos, NM 87545, USA}\\
{\small $^8$Columbia Astrophysics Laboratory, Columbia University, New York NY 10027, USA}\\
{\small $^9$Jet Propulsion Laboratory, California Institute of Technology, Pasadena, CA 91109, USA}\\
{\small $^{10}$Goddard Space Flight Center, Greenbelt, MD 20771, USA}\\
}

\begin{abstract}
We present the first images of the pulsar wind nebula (PWN) MSH 15$-$5{\sl2}
in the hard X-ray band ($\gapp$8~keV), as measured with
the {\em Nuclear Spectroscopic Telescope Array (NuSTAR)}.
Overall, the morphology of the PWN as measured by {\em NuSTAR} in the 3--7~keV band is similar to that
seen in {\em Chandra} high-resolution imaging. However, the spatial extent decreases with energy,
which we attribute to synchrotron energy losses as the particles move away from the shock.
The hard-band maps show a relative deficit of counts in the northern region towards the RCW~89
thermal remnant, with significant asymmetry.
\AR{We find that the integrated PWN spectra measured with {\em NuSTAR} and {\em Chandra}
suggest that there is a spectral break at 6~keV which may be explained by a break in the synchrotron-emitting
electron distribution at $\sim$200~TeV and/or imperfect cross calibration.}
We also measure spatially resolved spectra, showing that
the spectrum of the PWN softens away from the central pulsar B1509$-$58, and that there exists
a roughly sinusoidal variation of spectral hardness in the azimuthal direction. We discuss
the results using particle flow models.
We find non-monotonic structure in the variation with distance of spectral hardness
within 50$''$ of the pulsar moving in the jet direction,
which may imply particle and magnetic-field compression
by magnetic hoop stress as previously
suggested for this source. We also present 2-D maps of spectral parameters
and find an interesting shell-like structure in the $N_{\rm H}$ map.
We discuss possible origins of the shell-like structure and their implications.
\end{abstract}


\medskip
\section{Introduction}
A pulsar wind nebula (PWN) is a region of particles accelerated in a shock
formed by the interaction between the pulsar's particle/magnetic flux and
ambient matter such as a supernova remnant (SNR) or the interstellar medium (ISM).
It has been theorized that the shock, called a termination shock, can accelerate particles to
$\sim10^{15}\ \rm eV$, which are believed to contribute to the cosmic ray spectrum
from low energies up to the `knee' at $\sim10^{15}\ \rm eV$ \citep[e.g.,][]{dh92, aa96}.
In a PWN, the shock-accelerated particles propagate downstream and emit synchrotron photons under
the effects of the magnetic fields in that region \citep[][]{w72a, w72b, rg74}.
The electrons in the hard tail of the energy distribution produce X-rays,
and thus the detection of synchrotron
X-rays indirectly proves the existence of high energy electrons. Therefore, X-ray
emitting PWNe are particularly interesting for studying particle shock acceleration, and
for studying the interaction of high energy particles
with their environments \citep[see][for a review]{gs06}.

Since the accelerated particles in young PWNe lose their energy primarily via synchrotron radiation
at a rate proportional to $E^2$,
the energy distribution of the particles softens with distance from the shock,
an effect called `synchrotron burn-off'. As the particle spectrum softens, the emitted
photon spectrum is expected to soften as well. Details of the softening depend on the physical
environment and the particle flow in the PWNe; these have been modeled
using particle advection \citep[e.g.,][]{kc84, r03, r09} and/or diffusion \citep[e.g.,][]{g72,tc12}.
In particular, the advection models predict the radial profile of the photon index to be flat
out to the PWN edge and then to soften rapidly,
while diffusion models predict a gradual spectral softening with radius.
The particle spectrum can be inferred from the photon spectrum as they are directly related.
Therefore, spatially resolved spectra or energy resolved images can be compared to
model prediction
to infer the particle flow properties and the physical environments in PWNe.

MSH~15$-$5{\sl2} (also known as ``Hand of God'') is a large TeV-detected PWN which is powered
by the central 150-ms X-ray pulsar B1509$-$58 at a distance of $\sim$5.2 kpc \citep[][]{gbm+99}.
The radius of the PWN is $\sim 5'$ ($\sim 7.6$ pc) and it has a very asymmetric morphology
with complicated internal structures in the X-ray band \citep[e.g.,][]{gak+02, dga+06, yks+09}.
Notably, it has a hard jet directed south-east,
similar to that seen in some PWNe such as the Crab nebula \citep[][]{mbh+04}.
North of the PWN, there is a large thermal shell structure (RCW~89) which is thought
to be powered by the PWN through finger-like structures \citep[][]{ykk+05}.
The synchrotron burn-off effect in this PWN
was previously measured with {\em XMM-Newton} in the 0.5--10~keV band by \citet{sbd+10}.
They measured the burn-off effect in that band by integrating the spectrum azimuthally,
however, given the highly asymmetric morphology, a more in-depth analysis is warranted.

In this paper, we report on the spatial and spectral properties
of the PWN~MSH~15$-$5{\sl2} in a
broad X-ray band measured with {\em NuSTAR} and {\em Chandra}.
We present a two-dimensional synchrotron burn-off map
for the first time using energy-resolved images and spatially resolved spectra.
We describe the observations and data reduction in
Section~\ref{sec:obs}, and show the data analysis and results in Section~\ref{sec:ana}.
We discuss the implications of the analysis results in Section~\ref{sec:disc} and present the summary
in Section~\ref{sec:concl}.

\medskip
\section{Observations}
\label{sec:obs}
\newcommand{\markaa}{\tablenotemark{a}}
\begin{table}[t]
\vspace{0.0 mm}
\begin{center}
\caption{Summary of observations
\label{ta:obs}}
\scriptsize{
\begin{tabular}{cccccc} \hline\hline
Obs. No. & Observatory & Obs. ID & Exposure & Start Date \\
&	& 	&  (ks) & (MJD)&  \\ \hline
C1 & {\em Chandra} & 754  & 19 & 51770.6 \\ 
C2 & {\em Chandra} & 5534 & 49 & 53367.4 \\ 
C3 & {\em Chandra} & 5535 & 43 & 53408.6 \\ 
C4 & {\em Chandra} & 6116 & 47 & 53489.2 \\ 
C5 & {\em Chandra} & 6117 & 46 & 53661.0 \\ 
N1 & {\em NuSTAR} & 40024004002 & 42 & 56450.9 \\  
N2 & {\em NuSTAR} & 40024002001\markaa & 43 & 56451.8 \\  
N3 & {\em NuSTAR} & 40024003001 & 44 & 56452.6 \\  
N4 & {\em NuSTAR} & 40024001002 & 34 & 56519.6 \\ \hline 
\end{tabular}}
\end{center}
\vspace{0.0 mm}
\footnotesize{{\bf Notes.} All {\em Chandra} observations were made with the timed exposure mode (TE)
on ACIS-I chips.}\\
$^{\rm a}${ Only used for the spectral analysis along the jet and timing analysis.}\\
\vspace{-2.0 mm}
\end{table}

The {\em NuSTAR} instrument has two co-aligned hard X-ray optics and focal plane modules
(modules A and B with each module having four detectors), and
is the most sensitive satellite to date in the 3--79~keV band.
The energy resolution is 400~eV at 10~keV (FWHM), and the temporal resolution
is 2 $\mu$s \citep[see ][for more details]{hcc+13},
although the accuracy on orbital timescales is $\sim$2 ms due to
long-term clock drift. {\em NuSTAR} has unparalleled angular resolution in the
hard X-ray band (HPD=58$''$). The fine broadband angular response enables us to study
detailed morphological changes with energy for large PWNe such as MSH~15$-$5{\sl2},
and {\em NuSTAR}'s temporal resolution is sufficient to filter out contamination from the bright
central pulsar \citep[e.g., see ][for pulsars in PWNe]{c05, krh06}.

MSH~15$-$5{\sl2} was observed with {\em NuSTAR} in 2013 July with a total net exposure
of $\sim$160 ks. Although the {\em NuSTAR} field of view (FoV) is large enough to observe the whole PWN with a single pointing,
we used four different pointings in order to better sample different regions of the PWN.
We also analyzed archival {\em Chandra} ACIS \citep{gbf+03} observations in order to verify
our spatial analysis technique, and to broaden the energy range for spectroscopy
\citep[see also][]{gak+02,dga+06,yks+09}.
Table~\ref{ta:obs} summarizes the observations used in this paper.
The {\em NuSTAR} observation N2 was pointed to the jet, and most of the PWN
fell outside the FoV.
Therefore we used this observation only for the timing analysis in Section~\ref{sec:timingana}
and the spectral analysis along the jet direction in Section~\ref{sec:radspec}.
 
The {\em NuSTAR} data were processed with {\ttfamily nupipeline} 1.3.0 along with CALDB version 20131007, and
the {\em Chandra} data were reprocessed using the {\ttfamily chandra\_repro} tool of CIAO 4.5 along with
CALDB 4.5.7. We further processed the cleaned event files for analyses as described below.

\medskip
\section{Data Analysis and Results}
\label{sec:ana}

\medskip
\subsection{Timing Analysis}
\label{sec:timingana}
Since the pulsar B1509$-$58 is very bright, we needed to minimize its contamination in the PWN
imaging and spectral analysis. For the {\em Chandra} data, the pulsar contamination
was removed by image filtering. However, we were not able to do the image filtering for
{\em NuSTAR} because its point spread function (PSF) is broad.
Therefore, we selected the off-pulse interval in the {\em NuSTAR} data for the image and spectral
analyses below.

\begin{figure}
\centering
\vspace{-7.0 mm}
\hspace{-9.0 mm}
\includegraphics[width=3.7 in]{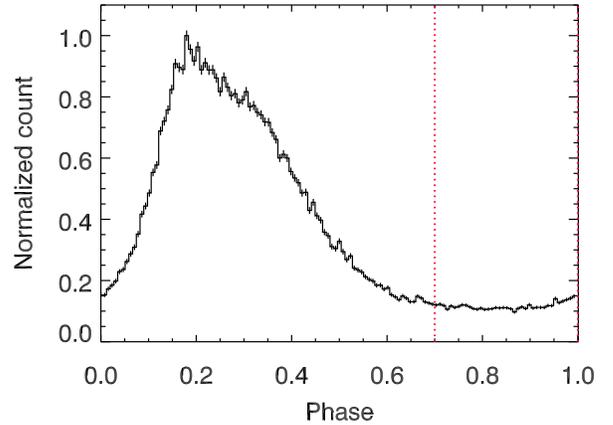} 
\vspace{-7.0 mm}
\figcaption{Normalized pulse profile for PSR~B1509$-$58 in the 3--79~keV band measured with {\em NuSTAR}.
Note that the off-pulse phases (0.7--1.0, dotted vertical lines) include DC emission
of the pulsar as well as the nebula emission.
\label{fig:PPdotprofiles}
}
\hspace{10.0 mm}
\end{figure}

We extracted the pulsar events in the {\em NuSTAR} observations in a $30''$ radius circle
in the 3--79~keV band, barycenter-corrected the events using R.A.=$15^{\rm h}13^{\rm m}55^{\rm s}.52$,
Decl.=$-59^{\circ}08'08\farcs$8 \citep[J2000,][]{cmb94} to produce event lists,
and divided each observation into three sub-intervals, yielding twelve sub-intervals for the
four observations.
We performed $H$ test \citep{dsr+89} on the event lists to measure the period for
the subintervals and fit the period to a linear function to find the spin period and the
spin-down rate. The pulsations were measured with very high significance in each subinterval, and
the measured period and the spin-down rate were $0.1517290191(14)$ s and
$1.5281(4)\times10^{-12}\ \rm s\ s^{-1}$ for 56450 MJD, respectively.
We folded the light curves using the measured period,
and show the resulting pulse profile in Figure~\ref{fig:PPdotprofiles}.
We used phases 0.7--1.0 for the PWN to minimize the pulsar contamination in all the
subsequent {\em NuSTAR} data analyses in this paper.
The other phase interval was used for the pulsar analysis which will be presented elsewhere.
We note that there is contamination from the DC emission of the pulsar even in the off-pulse interval.
For example, $\sim$1.6\% of the DC counts are expected in a circle of $R=30''$
at a distance of 60$''$ from the pulsar, but much less at larger distances.

\medskip
\subsection{Image Analysis}
\label{sec:imageana}

\begin{figure*}
\vspace{3.0 mm}
\hspace{0.0 mm}
\begin{tabular}{ccc}
\hspace{-3.0 mm}
\includegraphics[width=2.33 in]{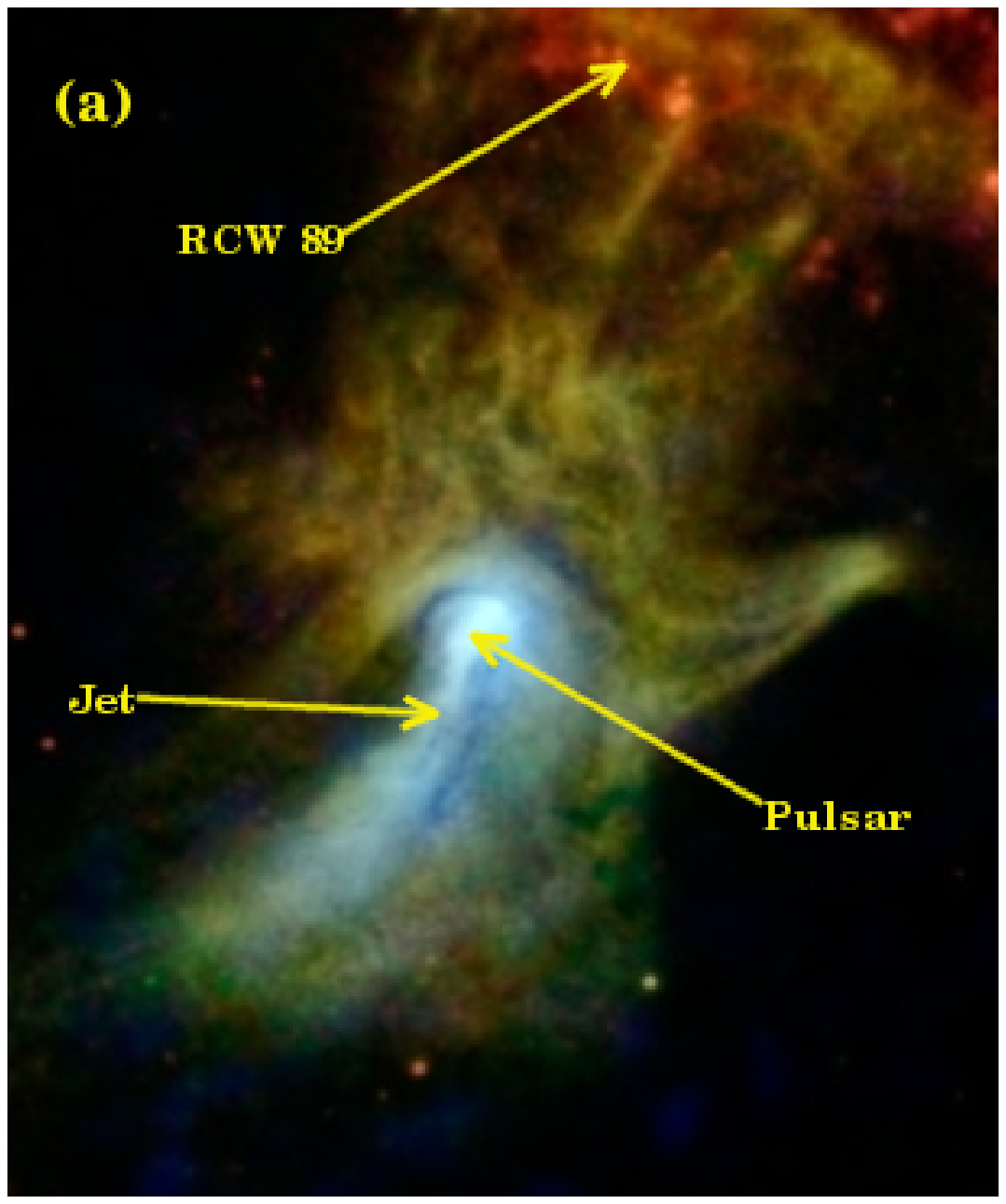} &
\hspace{-4.0 mm}
\includegraphics[width=2.33 in]{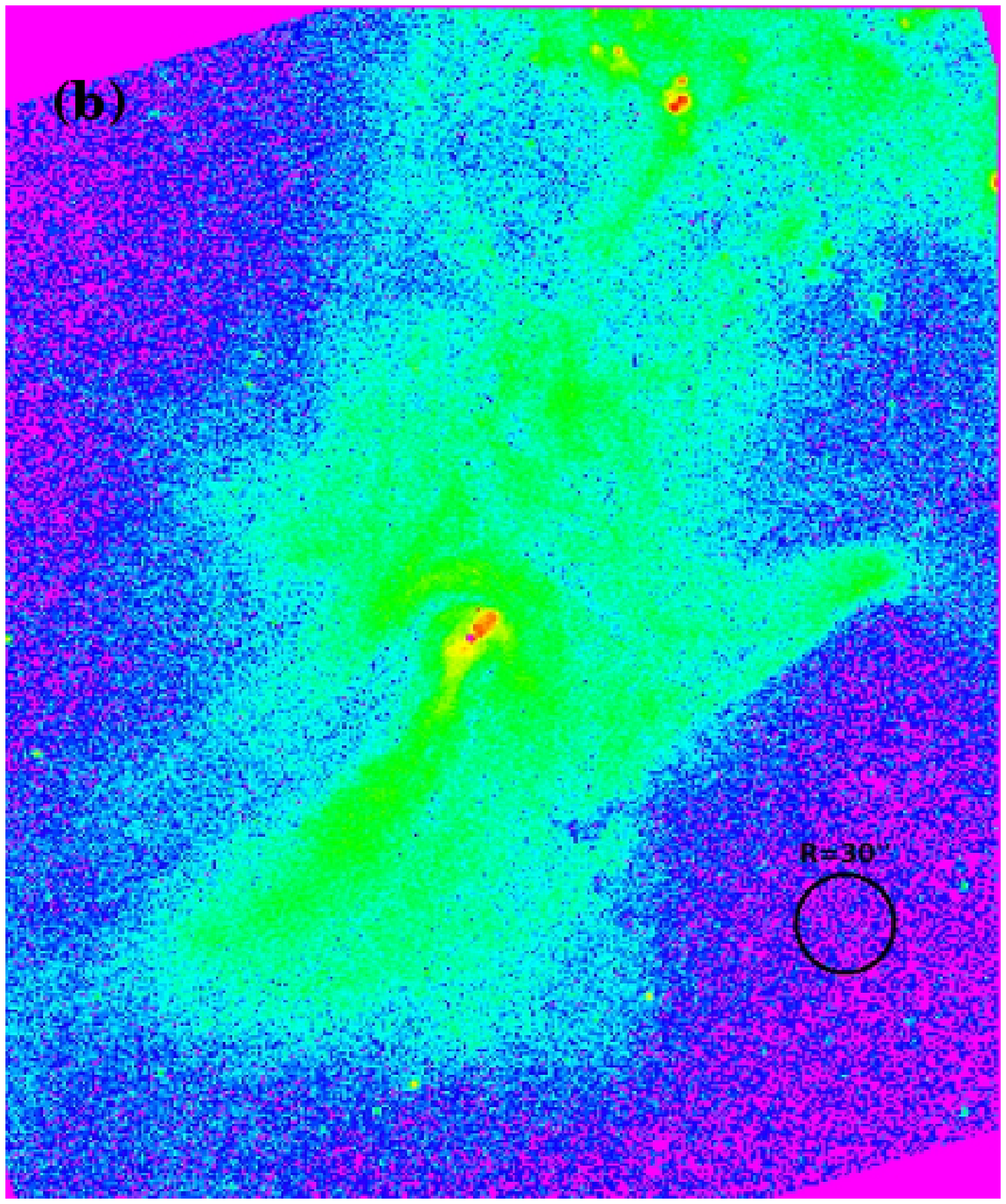} &
\hspace{-4.0 mm}
\includegraphics[width=2.33 in]{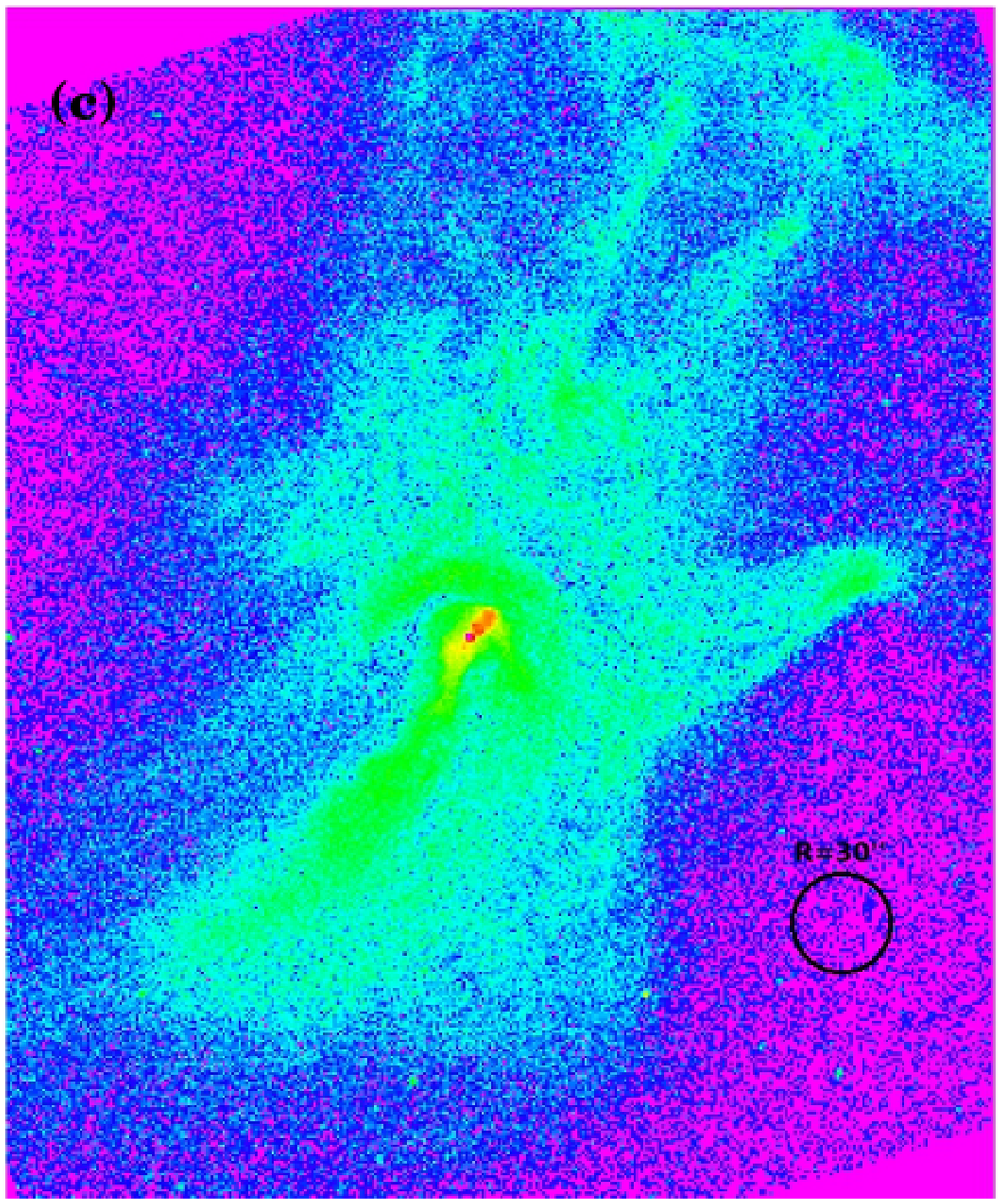} \\
\hspace{-3.0 mm}
\includegraphics[width=2.33 in]{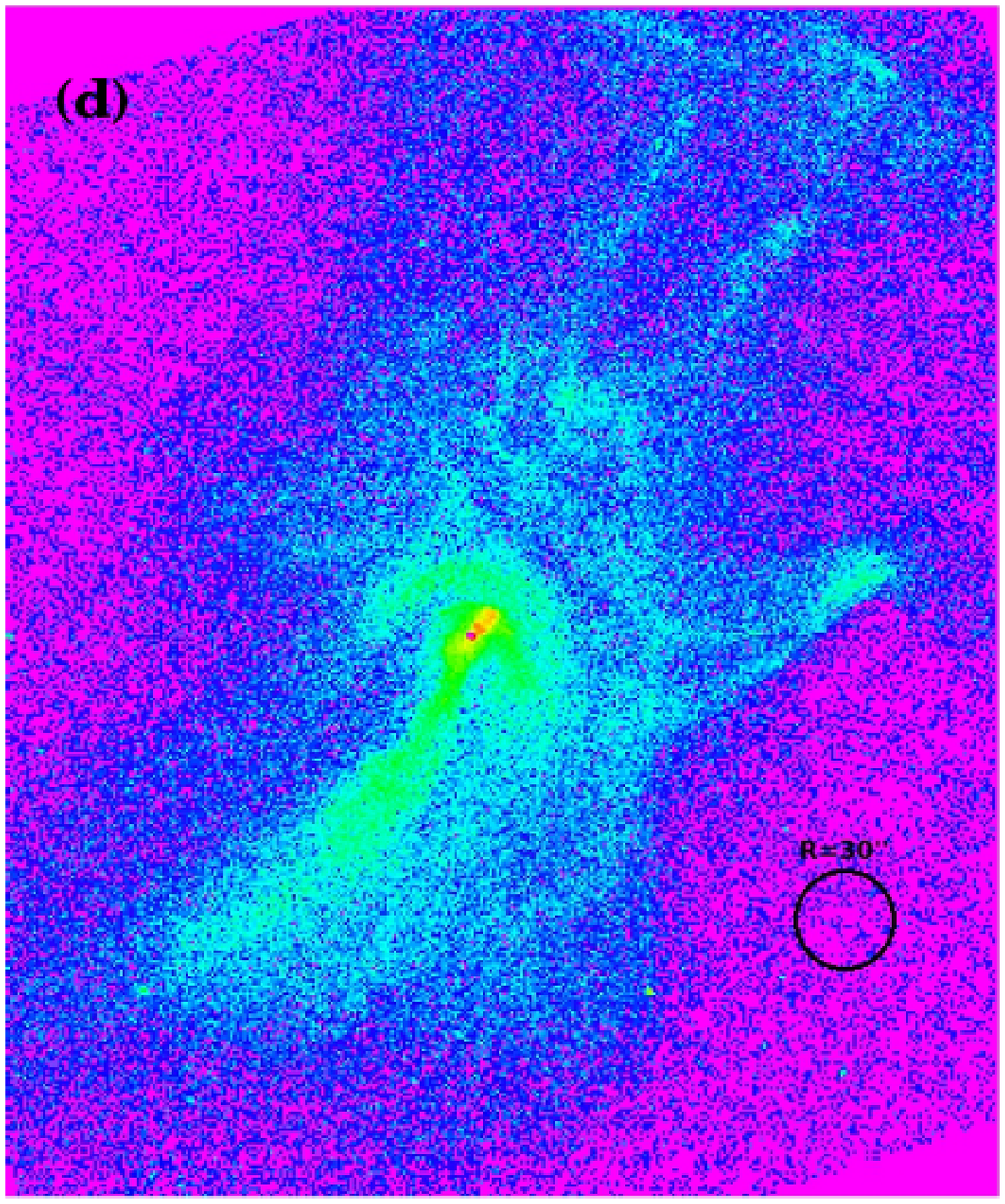} &
\hspace{-4.0 mm}
\includegraphics[width=2.33 in]{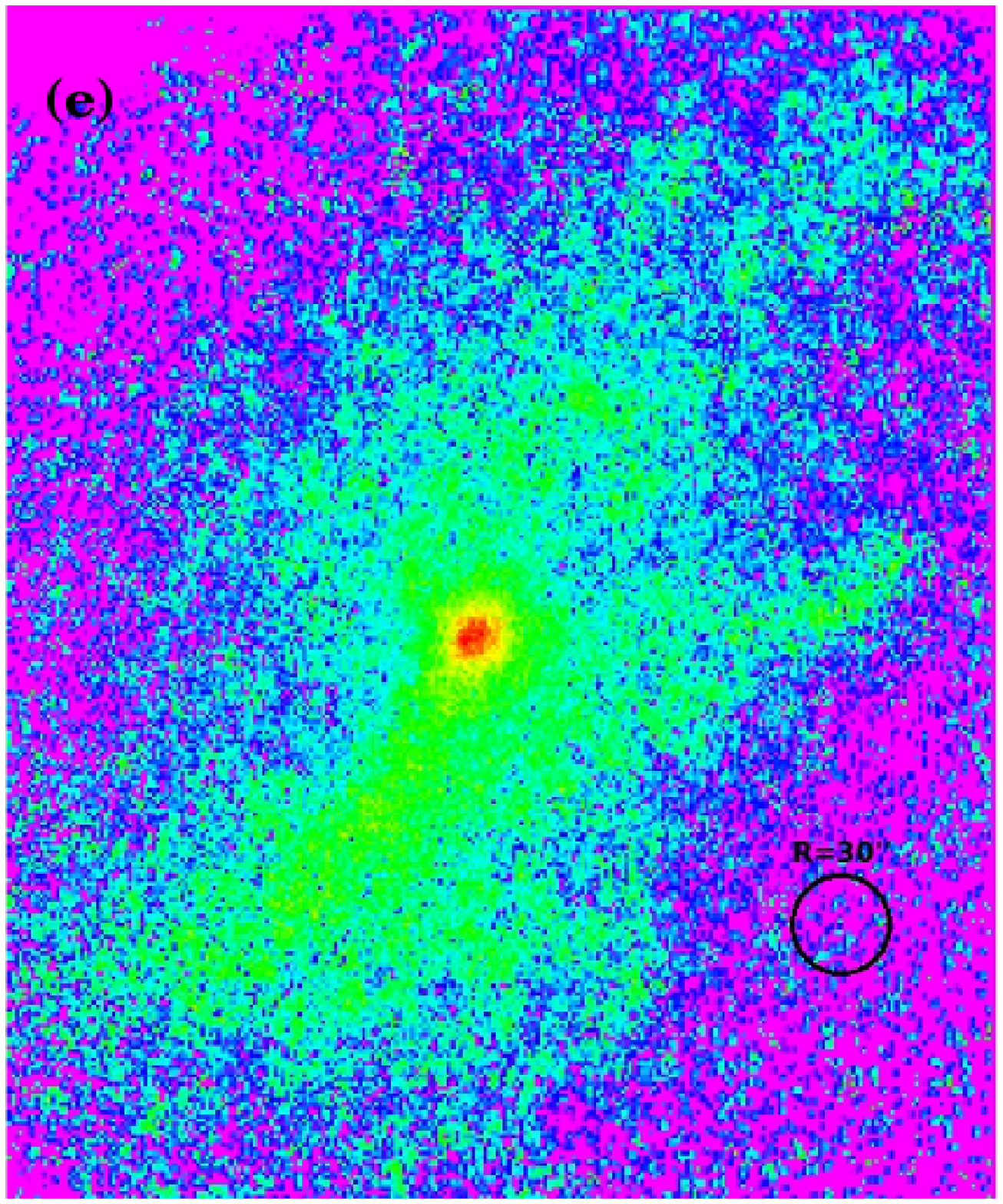} &
\hspace{-4.0 mm}
\includegraphics[width=2.33 in]{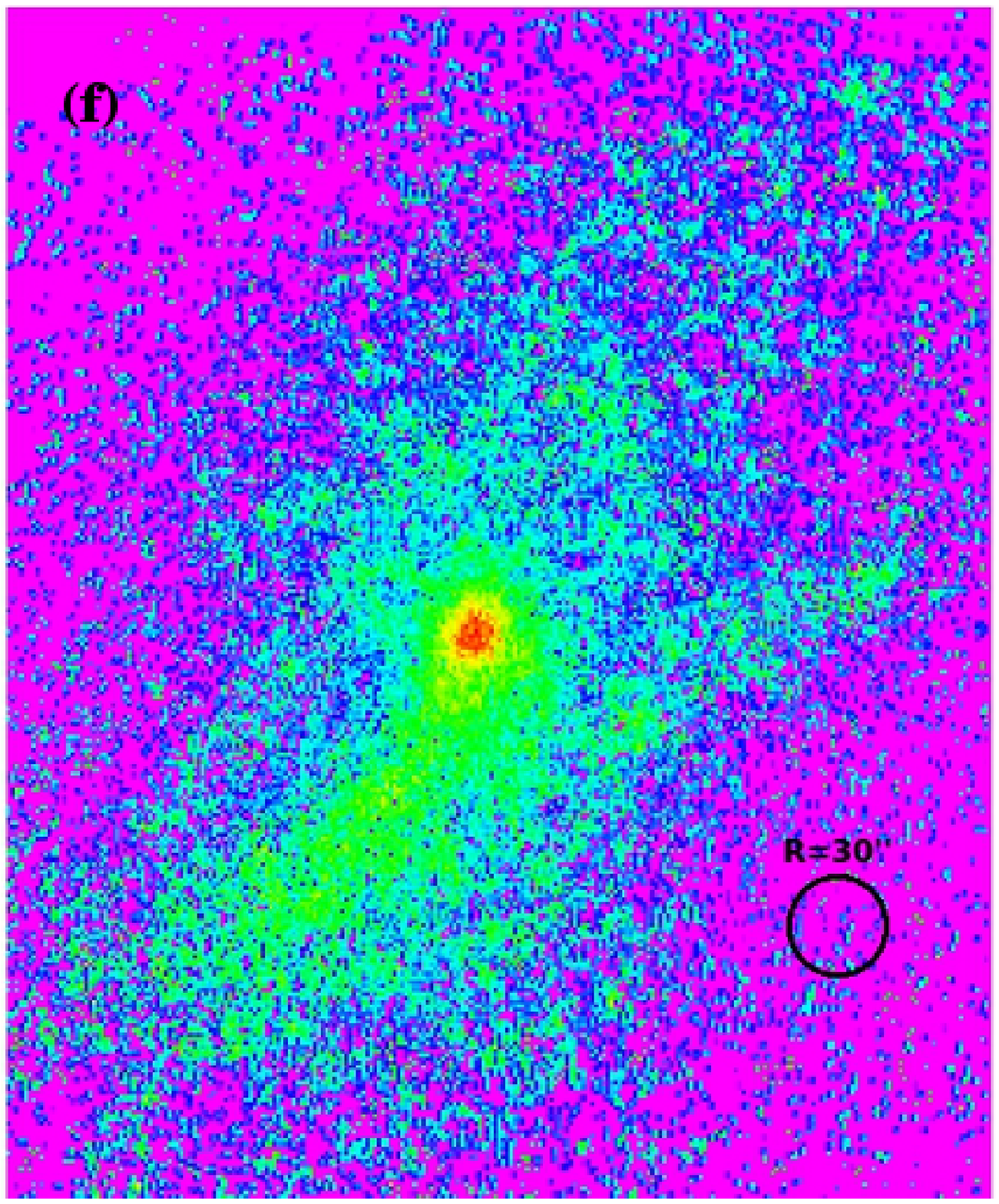} \\
\hspace{-3.0 mm}
\includegraphics[width=2.33 in]{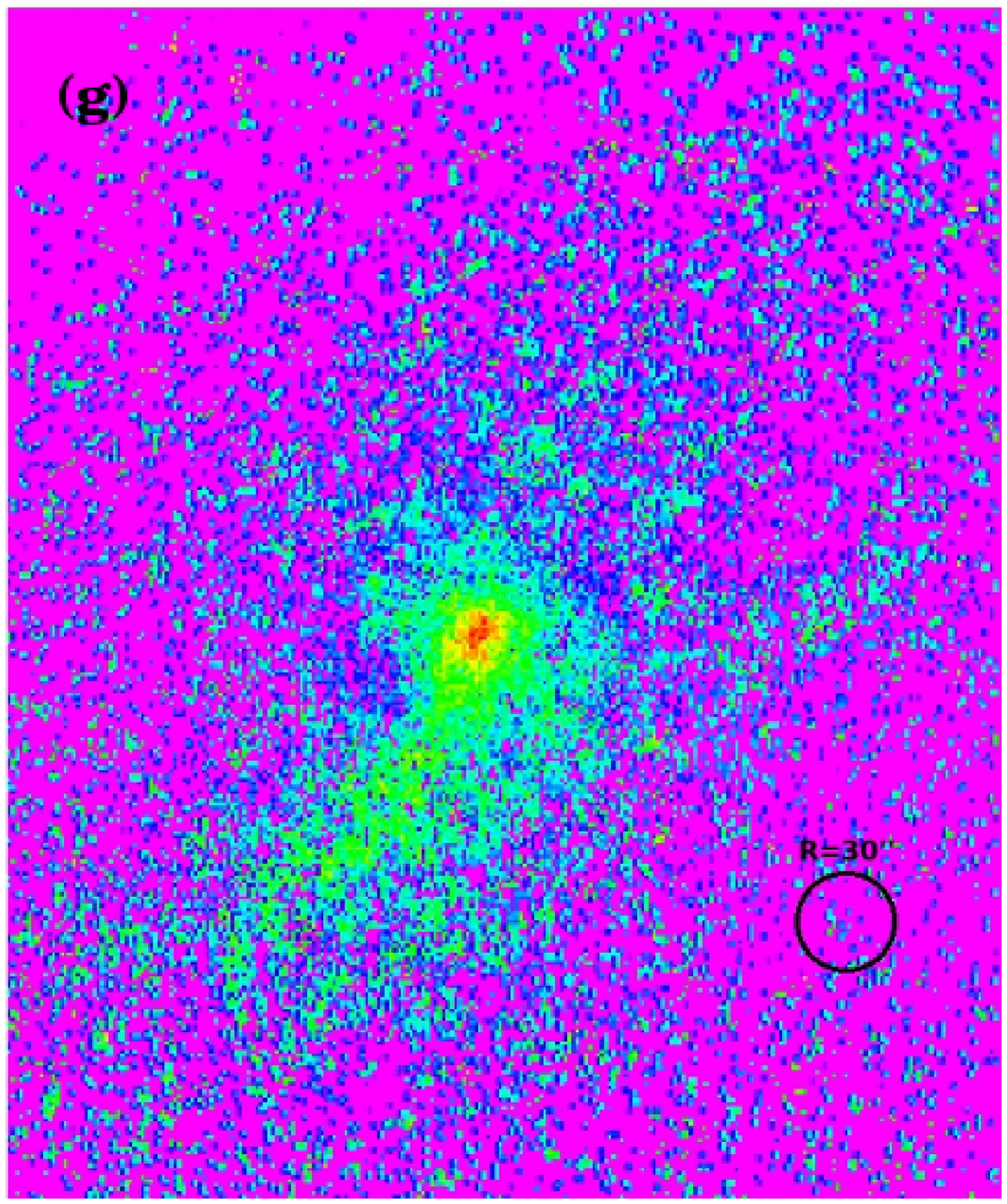} &
\hspace{-4.0 mm}
\includegraphics[width=2.33 in]{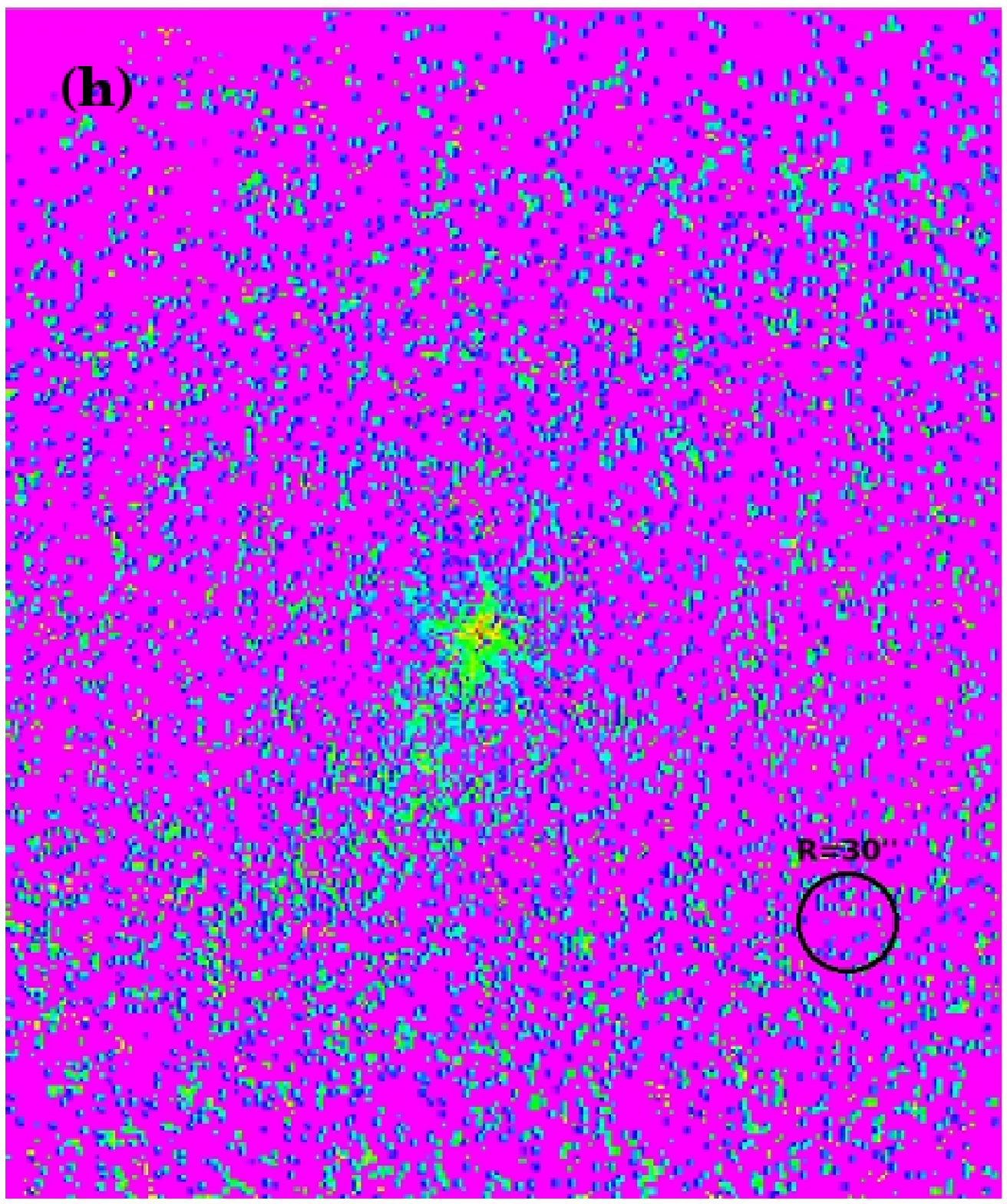} &
\hspace{-4.0 mm}
\includegraphics[width=2.33 in]{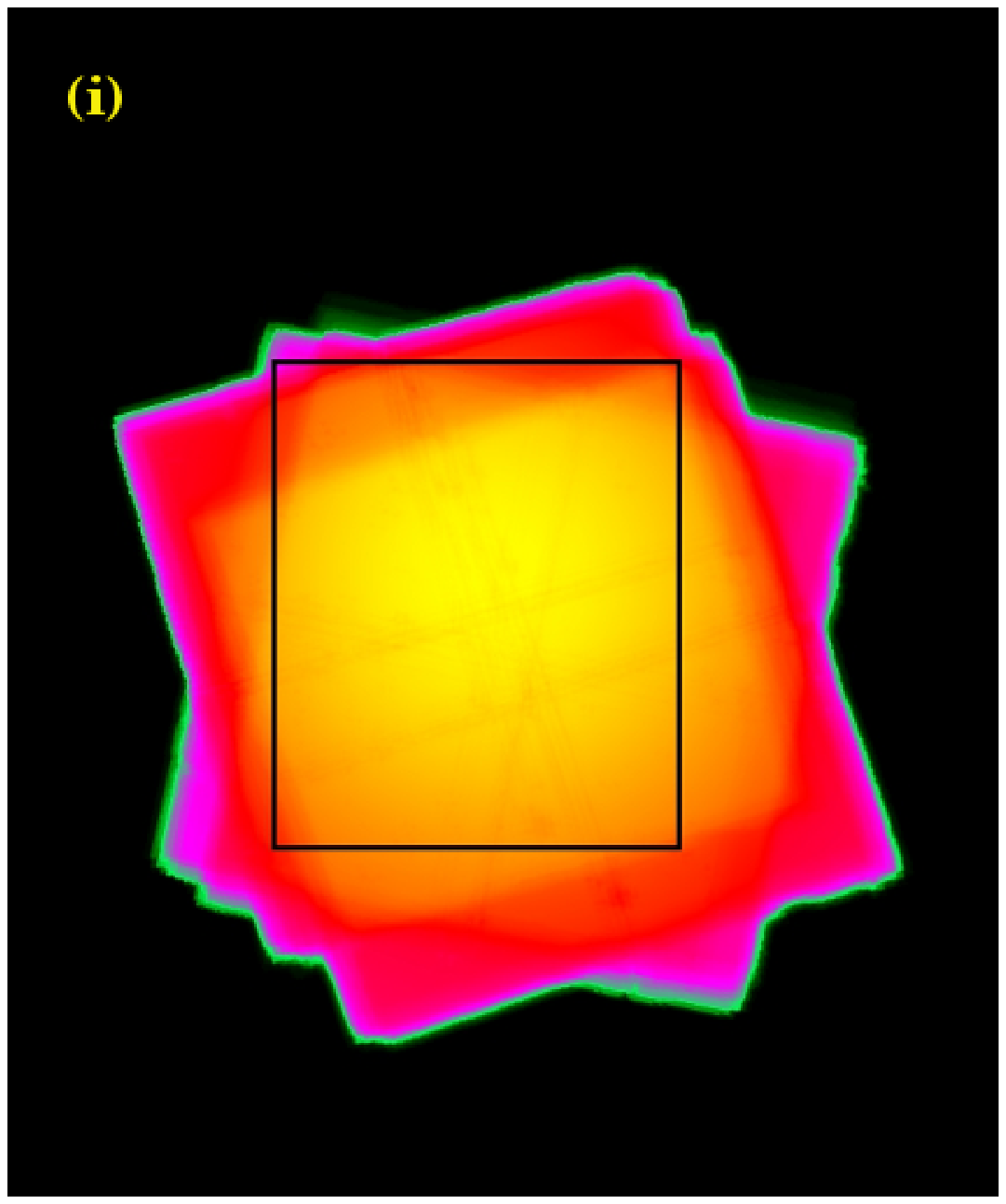} \\
\end{tabular}
\figcaption{MSH~15$-$5{\sl2} images measured with {\em NuSTAR} and {\em Chandra} in a $10'\times12'$ rectangular region:
({\it a}) A {\em NuSTAR} and {\em Chandra} combined false-color image in the 0.5--40~keV band
(see http://www.nustar.caltech.edu/image/nustar140109a), and intensity maps (b--h) of
({\it b}) {\em Chandra} 0.5--2~keV,
({\it c}) {\em Chandra} 2--4~keV,
({\it d}) {\em Chandra} 4--7~keV,
({\it e}) {\em NuSTAR} 3--7~keV,
({\it f}) {\em NuSTAR} 7--12~keV,
({\it g}) {\em NuSTAR} 12--25~keV,
({\it h}) {\em NuSTAR} 25--40~keV, and
({\it i}) {\em NuSTAR} exposure map and a box corresponding to the images.
For the {\em NuSTAR} data, we used off-pulse time intervals only in order to minimize the effect
of the central pulsar PSR~B1509$-$58. Exposure and vignetting corrections are applied to the images.
A circle with radius $30''$
is shown in panels b--h in black for reference. Note that the images use a logarithmic scale,
and each image has a different background level.
\label{fig:image}
}
\end{figure*}

In order to produce energy-resolved PWN images,
we first produced a merged {\em Chandra} image of the five observations in Table~\ref{ta:obs} using the
{\ttfamily merge\_obs} tool of CIAO 4.5 in the 0.5--2~keV, 2--4~keV and 4--7~keV bands
with bin size 4 pixels (Fig.~\ref{fig:image}). Note that the central 4$''\times$4$''$
corresponding to the pulsar emission was removed in these images.

For {\em NuSTAR} observations, we extracted events in the energy bands 3--7~keV, 7--12~keV,
12--25~keV, and 25--40~keV for the off-pulse phase.
After the phase selection, the pulsar component is expected to be reduced significantly.
The {\em NuSTAR} absolute aspect reconstruction accuracy
on long timescales is $\sim$8$''$ (90\% confidence),
which can blur the resulting merged image obtained with the three observations
and two modules. Therefore, we aligned the images
by registering the pulsar to the known position before phase filtering.
Since only one point source (the pulsar) was significantly detected in each observation, we were not able
to fully correct the position (e.g., for translation, rotation and scale).
We note that the rotational misalignments are measured and corrected with high accuracy \citep[][]{hlc+10},
and a small residual change in scale is not a concern for the spatial scales of our analyses.
Therefore, we assumed that the position offsets were caused by pure translations.

In order to produce deblurred images of the {\em NuSTAR} observations for comparison
with the low-energy high-resolution {\em Chandra} images,
we corrected for the exposure and deconvolved the {\em NuSTAR} images with the PSF using the
{\ttfamily arestore} tool of CIAO 4.5. We then
merged the deconvolved images (see Fig.~\ref{fig:image}).
The number of iterations in the deconvolution process was determined by
comparing the deconvolved {\em NuSTAR} image to the
{\em Chandra} image in a similar band. We chose the energy bands so that
the average photon energy weighted by the response and the spectrum in a {\em NuSTAR} band
is similar to that in a {\em Chandra} band, and used the 3--7~keV and 4--7~keV bands
for {\em NuSTAR} and {\em Chandra}, respectively.

\begin{figure}
\centering
\vspace{-6.0 mm}
\hspace{-10.0 mm}
\includegraphics[width=2.9 in, angle=90]{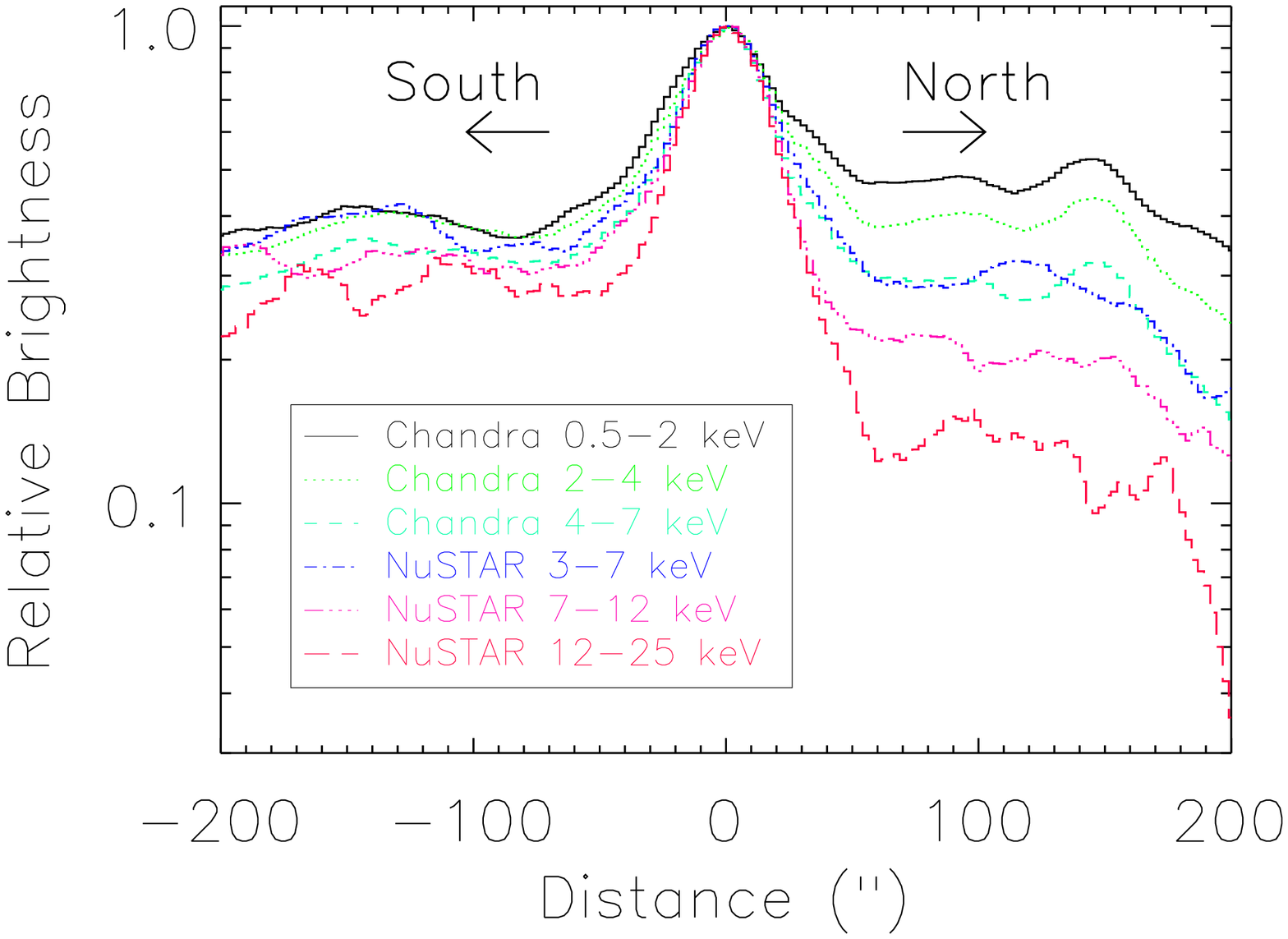} 
\vspace{-6.0 mm}
\figcaption{Projected profiles at several energy bands. The profiles are obtained by projecting
the images in Fig.~\ref{fig:image} onto the jet axis and smoothing over 25$''$.
\label{fig:comp}
}
\vspace{2.0 mm}
\end{figure}

Using the 2-D images, we produced projected profiles along the
jet axis (the south-east to north-west direction) in order to compare the deconvolved
3--7~keV {\em NuSTAR} profile with the 4--7~keV {\em Chandra} profile.
Here, we filled the pulsar region in the {\em Chandra} data which were
removed above with the average counts of the surrounding pixels.
We rotated the images 60$^{\circ}$ clockwise with the origin being the pulsar position,
so that the jet structure lies in the horizontal direction (x-axis).
We projected the images in Figure~\ref{fig:image} onto the axis along the jet, subtracted background,
smoothed the profile over a 25$''$ scale, and normalized the
scale with respect to the brightest point at the center.
The backgrounds were assumed to be flat over the detector chips. The background normalization factor
was first determined by taking a box in a source-free region, and then further adjusted by matching
the y-projected profiles of the source and the background at large distance from the center
for each energy band. We found that the results presented below are not sensitive to the background
subtraction since background accounts for only small fraction of the intensity.
We find that {\em NuSTAR}-measured profile in the 3--7~keV band is similar to that
measured with {\em Chandra} in the 4--7~keV band (see dashed and dot-dashed curves in fig.~\ref{fig:comp}),
and that the results of the deconvolution are not very sensitive to the number of iterations (e.g., 15--50),
and we used 20 iterations.

While the deconvolved {\em NuSTAR} image (Fig.~\ref{fig:image}e) shows similar overall morphology to
the high-resolution {\em Chandra} image in the similar band (Fig.~\ref{fig:image}d),
there are differences. Most notably,
the small arc-like structure and the elongation in the central region ($R\lapp30''$) are not
resolved in the {\em NuSTAR} images.
This is because the structures are smaller than the FWHM ($\sim$18$''$) of the {\em NuSTAR} PSF.
Also note that RCW~89, $\sim$6--7$'$ north, is not clearly visible in the {\em NuSTAR} data.
This is mainly because the {\em NuSTAR} observations did not have much exposure in that region;
most of RCW~89 fell outside the FoV during the observations.

\begin{figure*}
\vspace{0.0 mm}
\hspace{-10.0 mm}
\begin{tabular}{cc}
\includegraphics[width=4.0 in]{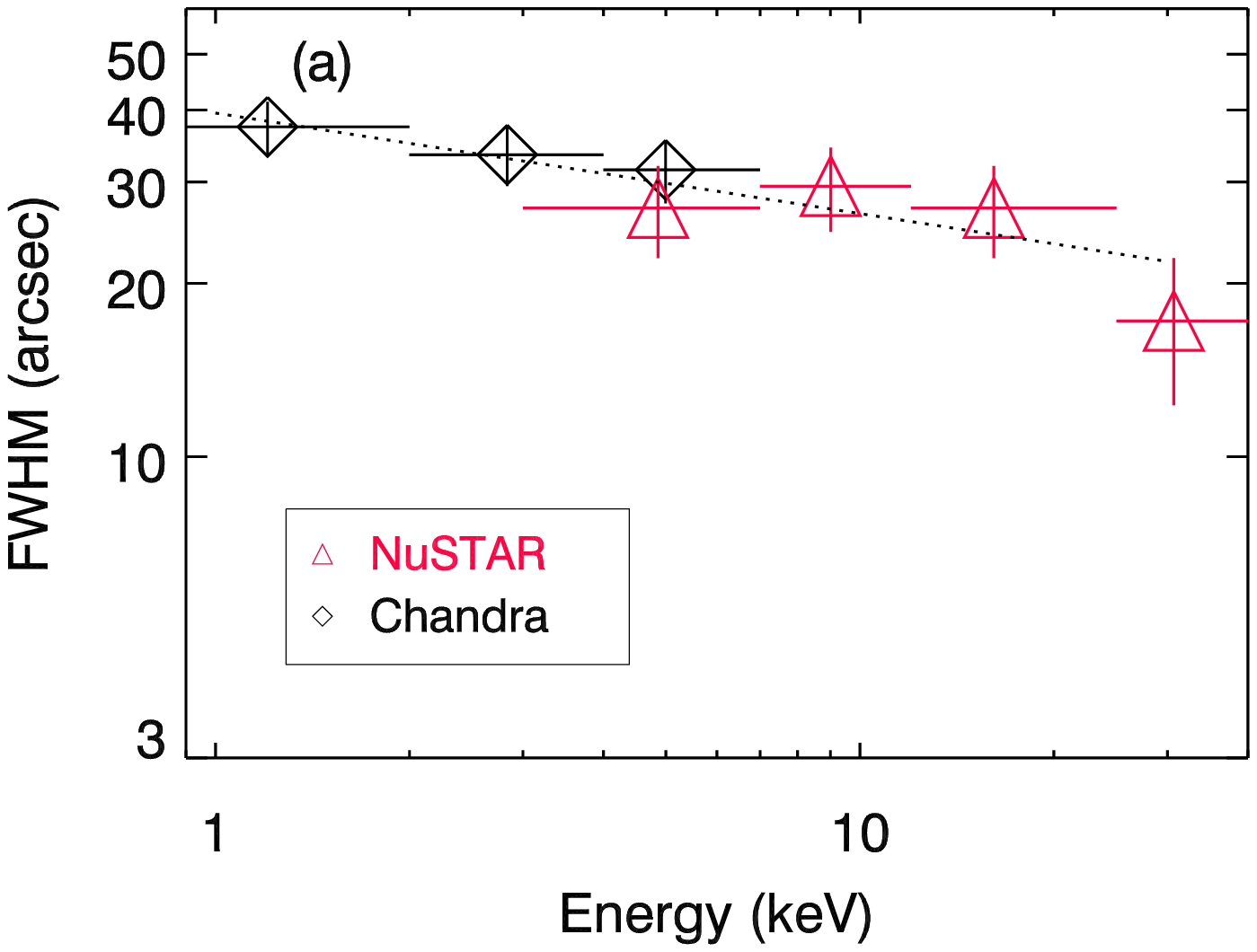} &
\hspace{-16.0 mm}
\vspace{-8.0 mm}
\includegraphics[width=4.0 in]{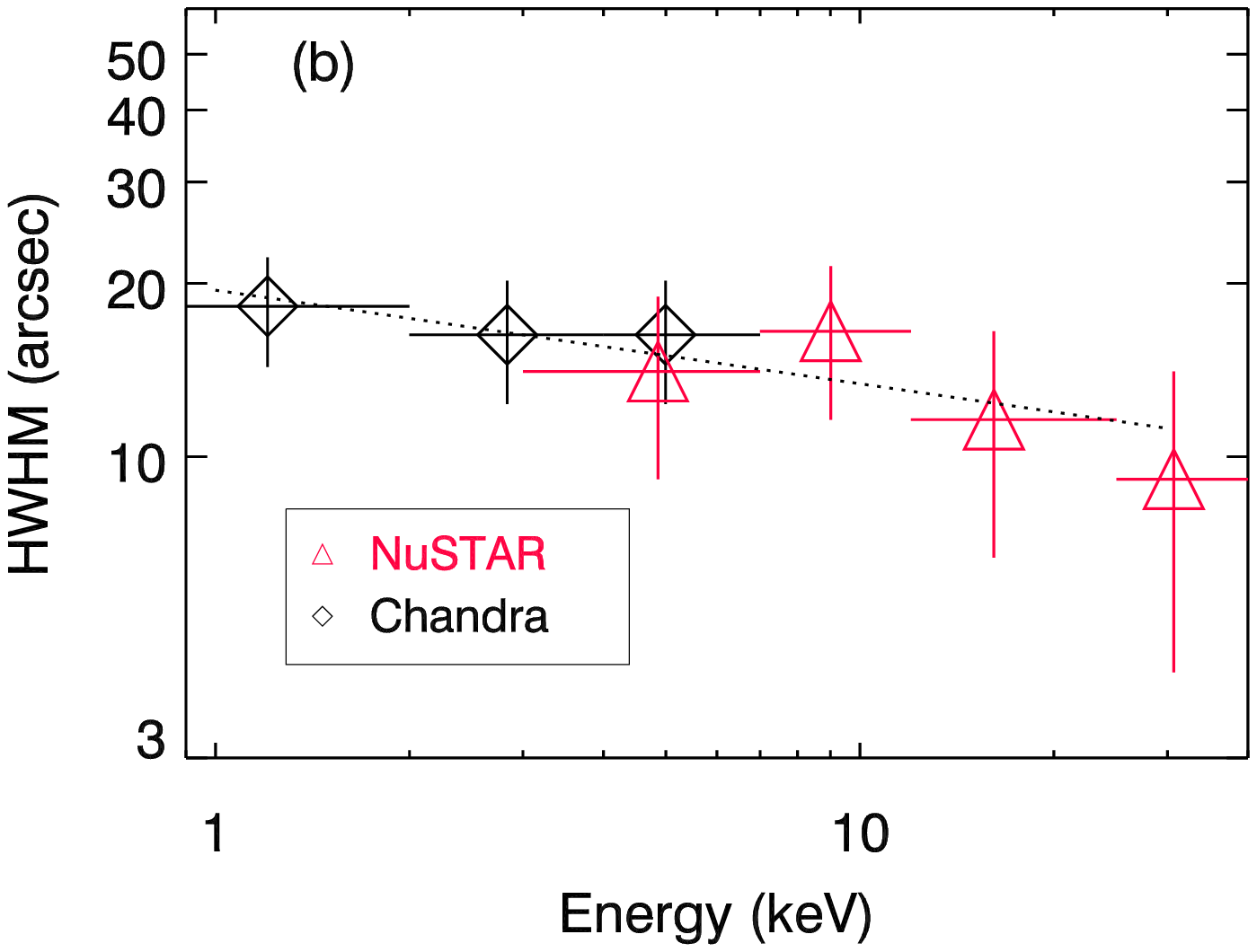} \\
\includegraphics[width=4.0 in]{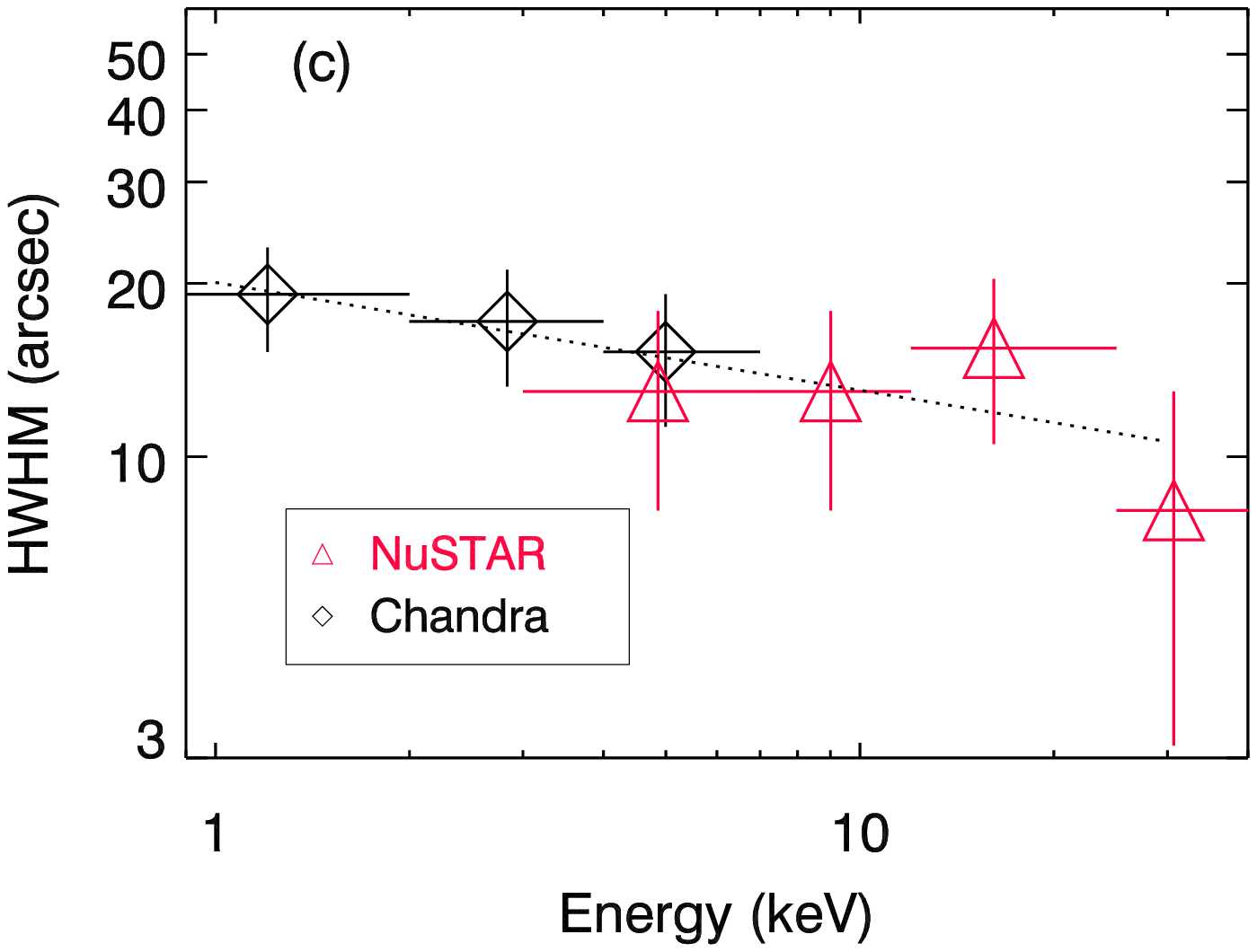} &
\hspace{-16.0 mm}
\includegraphics[width=4.0 in]{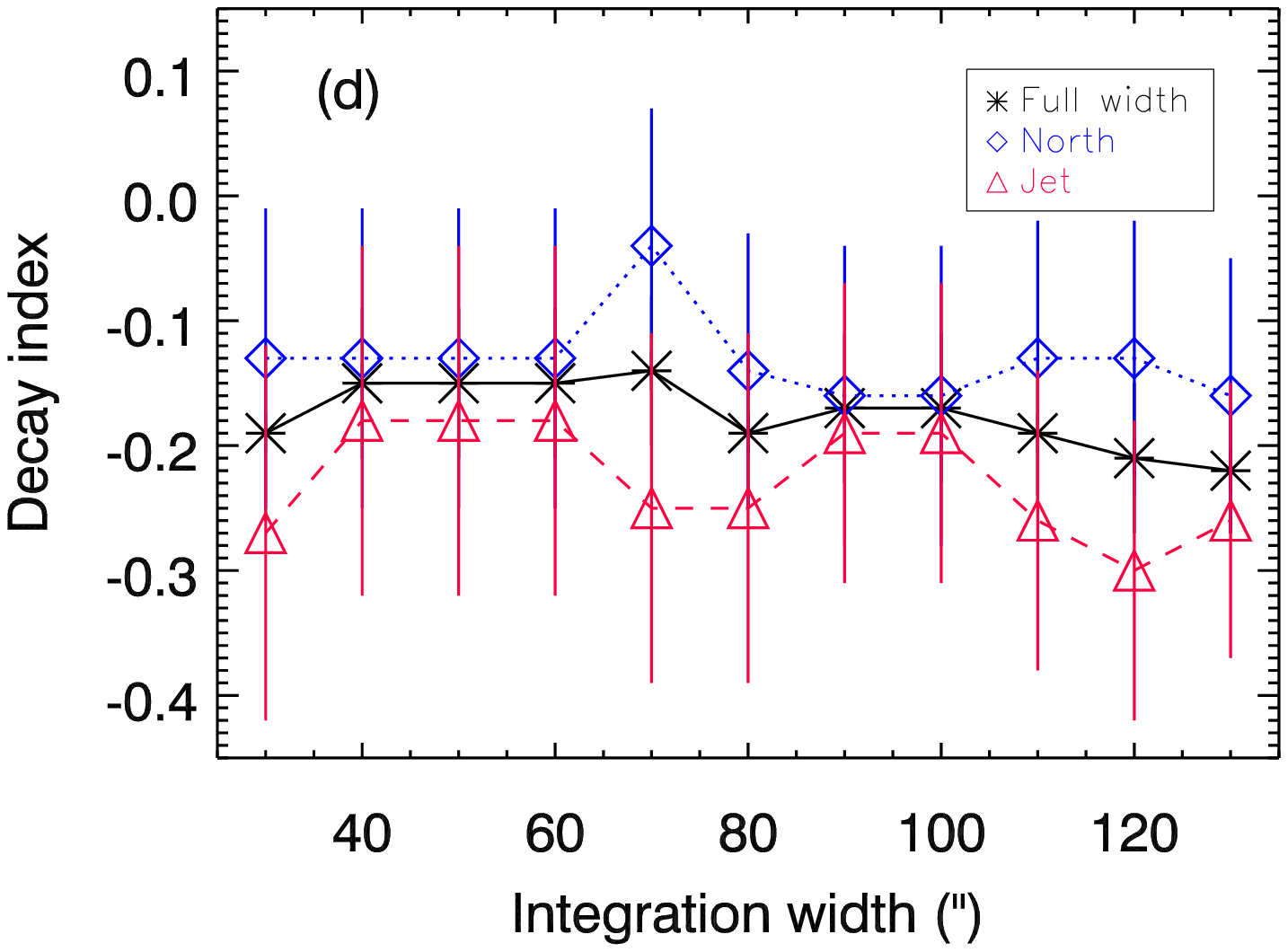} \\
\vspace{-6.0 mm}
\end{tabular}
\figcaption{Widths of projected profiles in several energy bands and
the best-fit power-law functions, $R(E)=R_{0} E^{m}$, for the FWHM (a),
HWHM of the northern (b) and the southern (c) nebulae for a width of 100$''$,
and the decay indices $m$ vs integration widths (d).
\label{fig:proj}
}
\end{figure*}

To measure the size of the narrow structures around the pulsar,
we projected events between $-50''$ and $50''$ in the y-axis direction onto the x-axis
in several energy bands.
The profile is very asymmetric and is not smooth on large scales ($R\gapp$100$''$).
Furthermore, the deconvolution produces artificial structures
in the outer regions due to the paucity of counts.
Therefore, defining a size (e.g., full width 1\% maximum) is
impractical on a large scale.
However, the source images are smooth on smaller scales ($R\lapp$50$''$),
allowing us to measure the FWHM and HWHM in the northern and the southern directions
of the projected profiles in several bands without smoothing the images.
We measured the sizes by calculating
the relative brightness with respect to the peak, and show them in Figures~\ref{fig:proj}a--c.
Although there are different structures in the northern and the southern directions,
the HWHM's are similar to each other and to half the FWHM.

We calculated the spectrum- and response-weighted average energy for each energy band,
fit the widths to a power-law function $R(E)=R_{0}E^{m}$ as suggested by \citet{r09},
and measured the decay index $m$ for various y-integration widths (e.g., $\sim$40--130$''$).
The measured decay index was stable over this range, as shown in Figure~\ref{fig:proj}d.
Note, however, that our measurement is based on deconvolved images,
and our uncertainties are therefore approximate.

\medskip
\subsection{Spectral Analysis}
\label{sec:spectrumana}
The image analysis shows a spectral change with radius in the PWN, and
we therefore tried to see differences in spectra at different radii.
We extracted spectra in various regions as described, and backgrounds from source-free regions,
and fit them with an absorbed power-law model.

Since spatial blurring due to the PSF size is much more significant
for {\em NuSTAR} than {\em Chandra}, we did not attempt to fit the spectrum jointly
except for one case of using a large aperture (Section~\ref{sec:totspec}), where
PSF ``blurring'' is not a large effect.
However, we jointly fit the spectra taken with single telescope at different epochs.
 
We used the $\chi^2$ and the {\ttfamily lstat} statistics
in {\ttfamily XSPEC} 12.8.1 to fit the spectra \citep[][]{a96}.
Results from the two methods were consistent,
and we primarily report the results obtained with the $\chi^2$ statistics.
Since the {\em NuSTAR} data are not sensitive to the hydrogen column density ($N_{\rm H}$) and
the results are not affected by small change of $N_{\rm H}$,
we froze it at a previously reported value \citep[$0.95\times 10^{22}\ \rm cm^{-2}$;][]{gak+02}.
We used a cross-normalization factor to account for a slight difference between {\em NuSTAR} module A and B,
and between observations. For the {\em Chandra} data fitting, we let $N_{\rm H}$ vary and introduced a
cross-normalization factor between observations.

\medskip
\subsubsection{Spectrum of the Entire Nebula}
\label{sec:totspec}

\newcommand{\marka}{\tablenotemark{a}}
\newcommand{\markb}{\tablenotemark{b}}
\newcommand{\markc}{\tablenotemark{c}}
\newcommand{\markd}{\tablenotemark{d}}
\begin{table*}[htb]
\vspace{0.0 mm}
\begin{center}
\caption{Best-fit parameters for the total PWN emission spectrum
\label{ta:spec}}
\scriptsize{
\begin{tabular}{ccccccccccccc} \hline\hline
Data\marka & Model\markb & Radius & Energy & $N_{\rm H}$\markc & $\Gamma_{\rm s}$ & $F_{\rm PL}$\markd & $E_{\rm break}$ & $\Gamma_{\rm h}$ & $\chi^2$/dof & \\ 
   &    &  $'$   & (keV) &($10^{22}\ \rm cm^{-2}$) &          &    & (keV) & &  &  &\\ \hline
C & PL      & 5  & 2--7   & 0.95  & 1.912(5) & 5.98(2)        & $\cdots$ & $\cdots$ & 1757/1704    \\
C & PL      & 5  & 3--7   & 0.95  & 1.90(1)  &  6.00(3)       & $\cdots$ & $\cdots$ & 1408/1359    \\ \hline
N & PL      & 5  & 3--7   & 0.95  & 2.03(2)  & 5.93(7)        & $\cdots$ & $\cdots$ & 543/587   \\ 
N & PL      & 5  & 3--20  & 0.95  & 2.06(1)  & 5.91(5)        & $\cdots$ & $\cdots$ & 1895/1887  \\ 
N & PL      & 5  & 15--30 & 0.95  & 2.1(1)   & 6.6(1)         & $\cdots$ & $\cdots$ & 576/579  \\ \hline
N + C & PL  & 5 & 2--20 & 0.95  & 1.950(4) & 5.77(5) & $\cdots$  & $\cdots$     & 3878/3592   \\ 
N + C & BPL & 5 & 2--20 & 0.95  & 1.918(5)  & 5.88(5) &  6.3(3)   & 2.12(2)      & 3691/3590   \\  \hline
\end{tabular}}
\end{center}
\vspace{-1 mm}
\footnotesize{{\bf Notes. $1\sigma$ uncertainties are given in parentheses
at the same decimal place as the last digit.} }\\
$^{\rm a}${ N: {\em NuSTAR}, C: {\em Chandra}.}\\
$^{\rm b}${ PL: {\ttfamily powerlaw} model,
and BPL: {\ttfamily bknpower} model in {\ttfamily XSPEC}.}\\
$^{\rm c}${ Frozen.}\\
$^{\rm d}${ Absorption corrected 3--10~keV flux in units of $10^{-11}\ \rm erg\ s^{-1} cm^{-2}$.}\\
\vspace{-2.0 mm}
\end{table*}

We first measured the total spectrum of the PWN using a source extraction aperture
of $R$=5$'$ centered at the pulsar position for a phase interval 0.7--1.0 in the {\em NuSTAR} data.
Photons with energies up to $\sim$20--30~keV were detected above background for each observation.
We extracted a spectrum for this aperture in each of the three {\em NuSTAR} observations,
N1, N3 and N4 in Table~\ref{ta:obs}, and jointly fit the spectra to a power law.
The measured power-law index is 2.06 and the absorption-corrected 3--10~keV flux is
5.9$\times 10^{-11}\ \rm erg\ cm^{-2}\ s^{-1}$ (see Table~\ref{ta:spec}).
For the {\em Chandra} data, we extracted source spectra using the same 5$'$ aperture,
ignoring the central 5$''$  in order to minimize the pulsar contamination,
and jointly fit the five {\em Chandra} spectra of each region to a power-law model
in the 0.5--7~keV band because background dominates above 7~keV.
We note that the {\em Chandra} data fits
were not acceptable with $\chi^2$/dof of 2632/2214 ($p=1\times 10^{-9}$),
having large residuals in the low energy band below 2~keV.
This is perhaps because the large regions are a mixture of subregions
with different spectra (e.g., see Section~\ref{sec:2dmap}).
We therefore fit the data above 2~keV only with frozen $N_{\rm H}$.
When removing photons below 2~keV, the remaining data were fit to a single power-law model with
a slightly smaller photon index, having $\chi^2$/dof=1757/1704 ($p=0.18$).
The cross-normalization factors for the five {\em Chandra} observations are all within 1\%.
Note that letting $N_{\rm H}$ vary also yields an acceptable fit ($\chi^2$/dof=1756/1703)
with $N_{\rm H}$=0.91(4) and $\Gamma=1.90(1)$.

The {\em Chandra}-measured spectrum in the 2--7~keV band is significantly harder
than that measured with {\em NuSTAR} in the 3--20~keV band.
We note that our results are consistent with the previous measurements made with {\em BeppoSAX}
and {\em INTEGRAL} \citep[][]{mcm+01,fhr+06}.
\citet{mcm+01} reported a photon index of 1.90(2) in the 1.6--10~keV band
for a 4$'$ aperture, which is consistent with our {\em Chandra} measurement in the 2--7~keV band.
In the hard band, the reported photon indices were 2.1(2) and 2.12(5) for {\em BeppoSAX} (20--200~keV)
and {\em INTEGRAL} (15--100~keV), respectively.
Note that the large apertures used for {\em BeppoSAX} and {\em INTEGRAL}
include the RCW~89 region, but the effect of RCW~89 is negligible
because the emission is very soft \citep[][]{ykk+05} and the telescopes operate only above 15~keV.
The photon index we measure with {\em NuSTAR} in the 15--30~keV band is 2.1(1)
for the 5$'$-aperture, which agrees with the previous measurements.
The results of our measurements are summarized in Table~\ref{ta:spec}. 

Since the large apertures include many subregions with different spectral properties as we show
below (see Sections~\ref{sec:azspec}, \ref{sec:radspec}, and \ref{sec:2dmap}), a single power
law may not properly represent the combined spectrum. In particular, we find that
the best-fit photon index for the {\em Chandra} data becomes smaller
as we ignore lower energy spectral channels, that is, the spectrum appears to
harden (is concave up) as we move to higher energies.
However, this is the opposite to what we see with {\em NuSTAR} (Table~\ref{ta:spec}).
While this may imply a spectral break in the X-ray band, some other effects such as
contamination from the pulsar and/or RCW~89\AR{, or cross-calibration systematics between the two instruments}
may have some impact.
Therefore, we investigate some possibilities below in order to see if the discrepancy in the spectral
index measurements of {\em NuSTAR} and {\em Chandra} is caused by a spectral break.

\begin{figure*}[htb]
\vspace{0.0 mm}
\hspace{-3.0 mm}
\begin{tabular}{cc}
\includegraphics[width=2.5 in, angle=90]{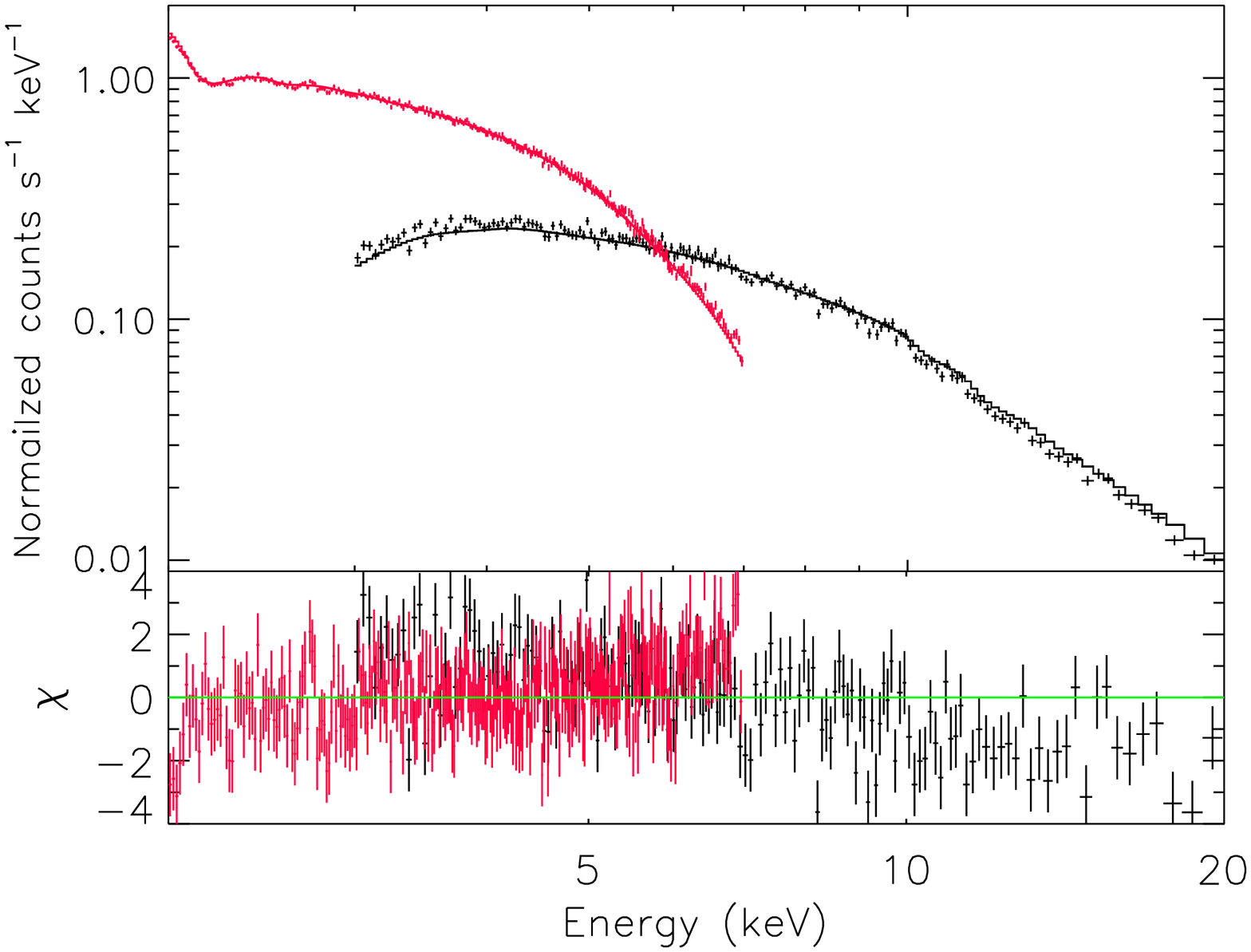} &
\hspace{7.0 mm}
\includegraphics[width=2.5 in, angle=90]{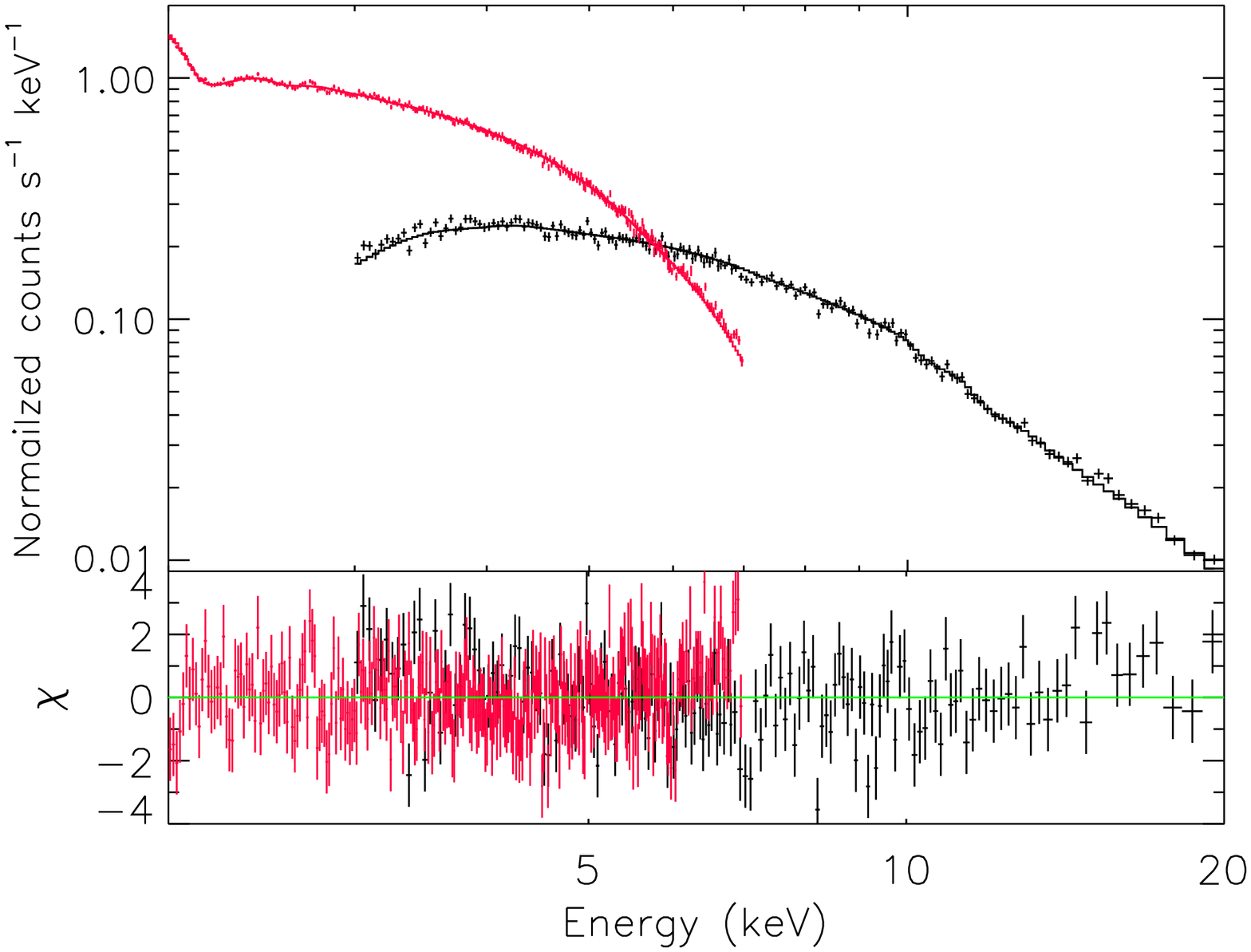} \\
\end{tabular}
\figcaption{Joint fit results of the {\em NuSTAR} and the {\em Chandra} spectra for
the single power-law (left) and the broken power-law (right) models.
Spectra obtained with each telescope are merged for display purpose only.
\label{fig:spec}
}
\end{figure*}

First, we note that the pulsar contamination was not completely removed
in the {\em Chandra} data. Excising 5$''$ leaves 2--5\% pulsar emission
in the $R=5'$ aperture, which may bias the PWN spectrum.
In order to see the effects quantitatively, we fit the pulsed spectrum
of the pulsar in the {\em NuSTAR} data
(total spectrum minus the DC level in Fig.~\ref{fig:PPdotprofiles})
with a power-law model and find that the photon index is 1.36(1)
and the 3--10~keV flux is
$2.24(3)\times 10^{-11}\ \rm erg\ s^{-1}\ cm^{-2}$ which broadly agree with the previous
measurements \citep[][]{cmm+01,glq+12}. We added this pulsar component to the {\em Chandra} fit
assuming 5\% of the pulsar emission is in the 5$'$ aperture after the 5$''$ excision.
Thus, the spectral model was a double power-law model, one for the pulsar emission
and the other for the PWN emission. We froze the pulsar component, fit the PWN spectrum, and
find that the spectral index of the PWN does not change. Since the spectral index of
the pulsed spectrum may be different in the soft band, we changed the spectral index of the pulsar
component to 1.19 as reported by \citet{cmm+01} in the 1.6--10~keV band, and find that the
photon index of the entire PWN softens only by $\Delta \Gamma=0.01$. We further increased
the pulsar flux by 10\% and find no change in the spectral index of the PWN. We verified the results
by increasing the excision region to 10$''$.
Note that the DC component of the pulsar is
not included in this study. However, the unmodeled DC component is much smaller than the pulsed
component as seen in Figure~\ref{fig:PPdotprofiles}, so the effect would be negligible
in the {\em Chandra} data fit.

Second, we consider the effect of the pulsar DC component in the {\em NuSTAR} data.
Although the DC component is negligible in the {\em Chandra} data due to the image filtering,
the DC component presents in the {\em NuSTAR} data because time filtering does not remove the DC emission.
Although it is not possible to measure the DC spectrum accurately,
we estimated 3--10~keV DC flux using a 30$''$ aperture as follows.
With {\em NuSTAR}, we measure the total (pulsed+DC+nebula) and the pulsed flux to be
$3.29\times10^{-11}\ \rm erg\ s^{-1}\ cm^{-2}$ and $2.24\times10^{-11}\ \rm erg\ s^{-1}\ cm^{-2}$, respectively.
The nebula flux is measured to be $7.5\times10^{-12}\ \rm erg\ s^{-1}\ cm^{-2}$ with {\em Chandra}
(Section~\ref{sec:radspec}). By subtracting the pulsed and the nebula flux from the total flux, the 3--10~keV
DC flux is estimated to be $3.1\times10^{-12}\ \rm erg\ s^{-1}\ cm^{-2}$. We assumed that the photon index is
1.7, similar to the {\em NuSTAR}-measured value for the 30$''$ aperture (Section~\ref{sec:radspec}).
We included the DC emission in the {\em NuSTAR} fit of the $5'$-aperture spectrum,
and followed the procedure described above for the pulsar contamination estimation
in the {\em Chandra} data. This procedure effectively removes the DC component from the PWN spectrum.
However, note that removing
such a hard spectrum only softens the spectrum of the entire PWN, making the discrepancy larger.
We therefore arbitrarily changed the photon index of the DC component to 2.5 to mitigate the possibility
of having very soft DC emission and find that the photon index of the entire PWN hardens only by $\Delta \Gamma=0.02$.

Third, we varied the background level by $\pm$30\%,
and found that spectral indices change only by 0.02 and 0.01 for
{\em Chandra} and {\em NuSTAR}, respectively. We also used different background regions and found that
the spectral indices do not change significantly.

Finally, we estimate the contamination of the RCW~89 emission in the {\em NuSTAR} data.
Using the {\em NuSTAR} PSF, we estimated the contamination of a structure
at $\sim$6$'$ into the $5'$ circle to be $\sim$14\%.
We extracted the RCW~89 spectrum from the {\em NuSTAR} observation N1 which sampled
the RCW~89 best among the {\em NuSTAR} observations. We added the spectrum as additional
background in the PWN spectral fit.
Since the other observations N3 and N4 sampled only a small fraction of
the RCW~89 region, we used the RCW~89 spectrum extracted from N1 for these observations as well.
We varied the normalization of the RCW~89 background from 0.14 to 0.28 in order to account for the fact
that the actual RCW~89 region may be larger than what we sampled with {\em NuSTAR},
and found that the photon index changes by $\lapp$0.02.

The large discrepancy in the spectral index measurements between
{\em NuSTAR} and {\em Chandra} cannot be explained by a combination of the above effects.
\AR{We therefore consider alternatives below. We note that there may be cross-calibration
systematics between {\em NuSTAR} and {\em Chandra}. For example, \citet{kbb+05} showed that
systematic uncertainties
between X-ray observatories caused by cross calibration are significant using observations
of the Crab nebula, for which the authors found that {\em Chandra} and {\em BeppoSAX}/LECS
measured smaller spectral indices than {\em BeppoSAX}/MECS and {\em INTEGRAL} did.
Since {\em NuSTAR} is calibrated so that the spectrum of the Crab nebula is a simple power
law with a photon index of 2.1, larger than the {\em Chandra}-measured value of 1.95 \citep{kbb+05},
the spectral break we see for MSH~15$-$5{\sl 2} could be explained by imperfect cross-calibration.}

\AR{However, the cross-calibration effects may be different from source to source depending on the spectral shape,
and the measurements made for the Crab nebula may not be directly translated into the case of MSH~15$-$5{\sl 2}.}
We therefore consider \AR{an} alternative; the discrepancy of spectral indices between the {\em Chandra}
and the {\em NuSTAR} measurements is caused by a spectral break.
We jointly fit the 2--7~keV {\em Chandra} and the 3--20~keV {\em NuSTAR} data
with a single power law and a broken power law. We find that a single power law
does not describe the data well having $\chi^2$/dof=3878/3592 ($p=5\times10^{-4}$),
but a broken power law with a break energy of 6.3~keV
does ($\chi^2$/dof=3691/3590, $p=0.12$).
The results are presented in Table~\ref{ta:spec} and Figure~\ref{fig:spec}.

\label{sec:azspec}
\begin{figure*}[ht]
\centering
\begin{tabular}{ccc}
\hspace{-3.0 mm}
\vspace{0.0 mm}
\includegraphics[width=2.45 in]{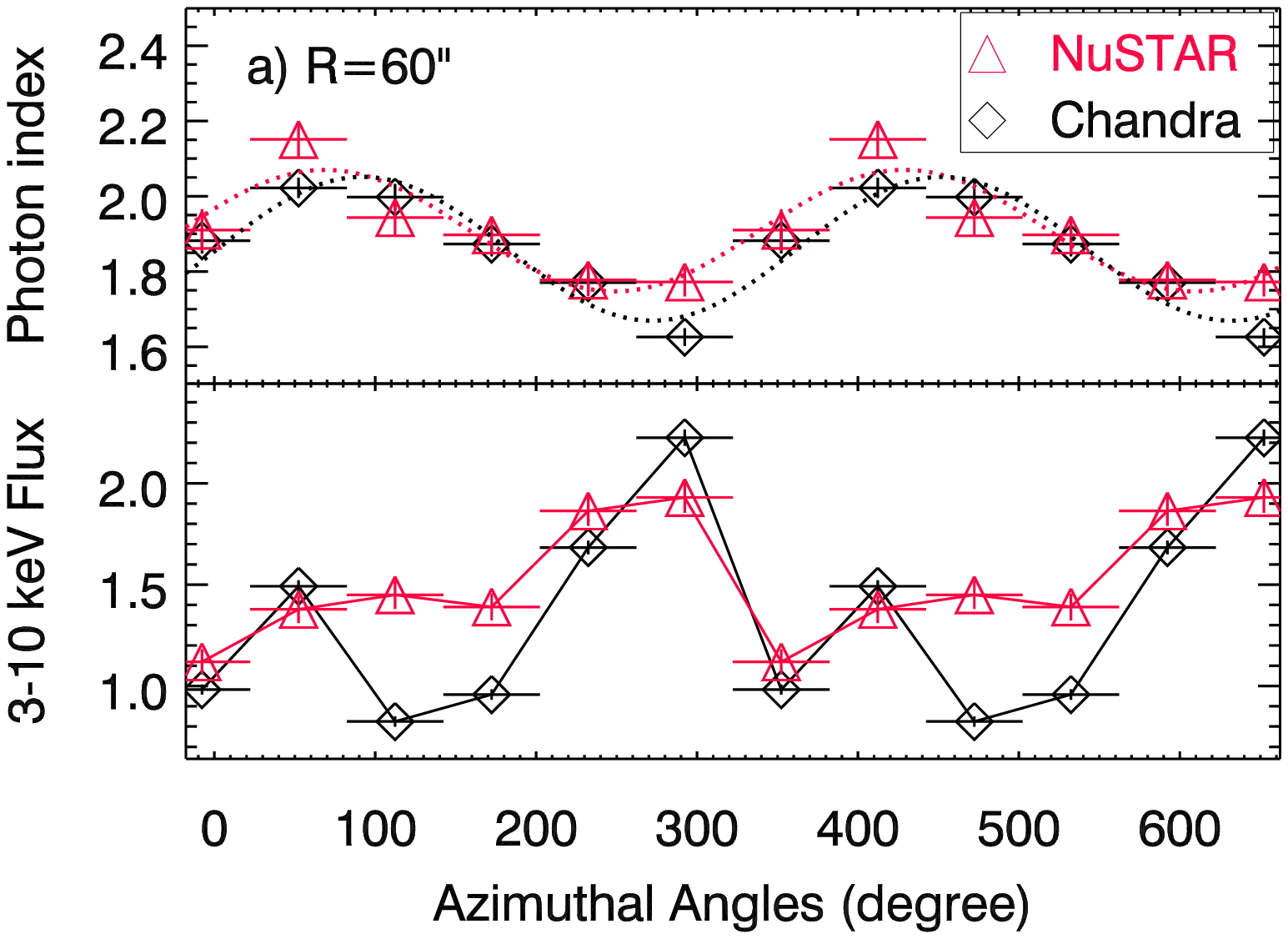} &
\hspace{-8.0 mm}
\includegraphics[width=2.45 in]{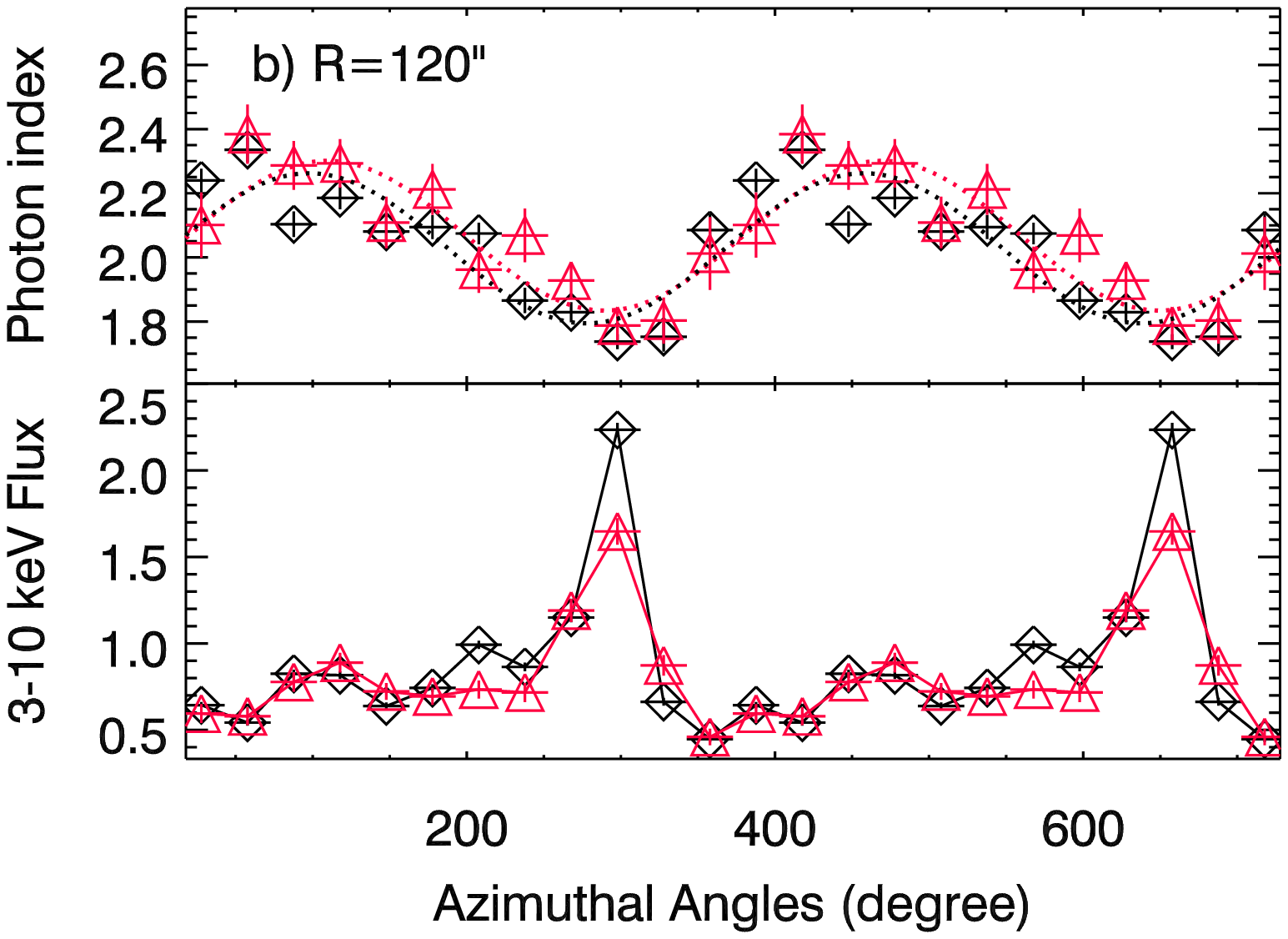} &
\hspace{-8.0 mm}
\includegraphics[width=2.45 in]{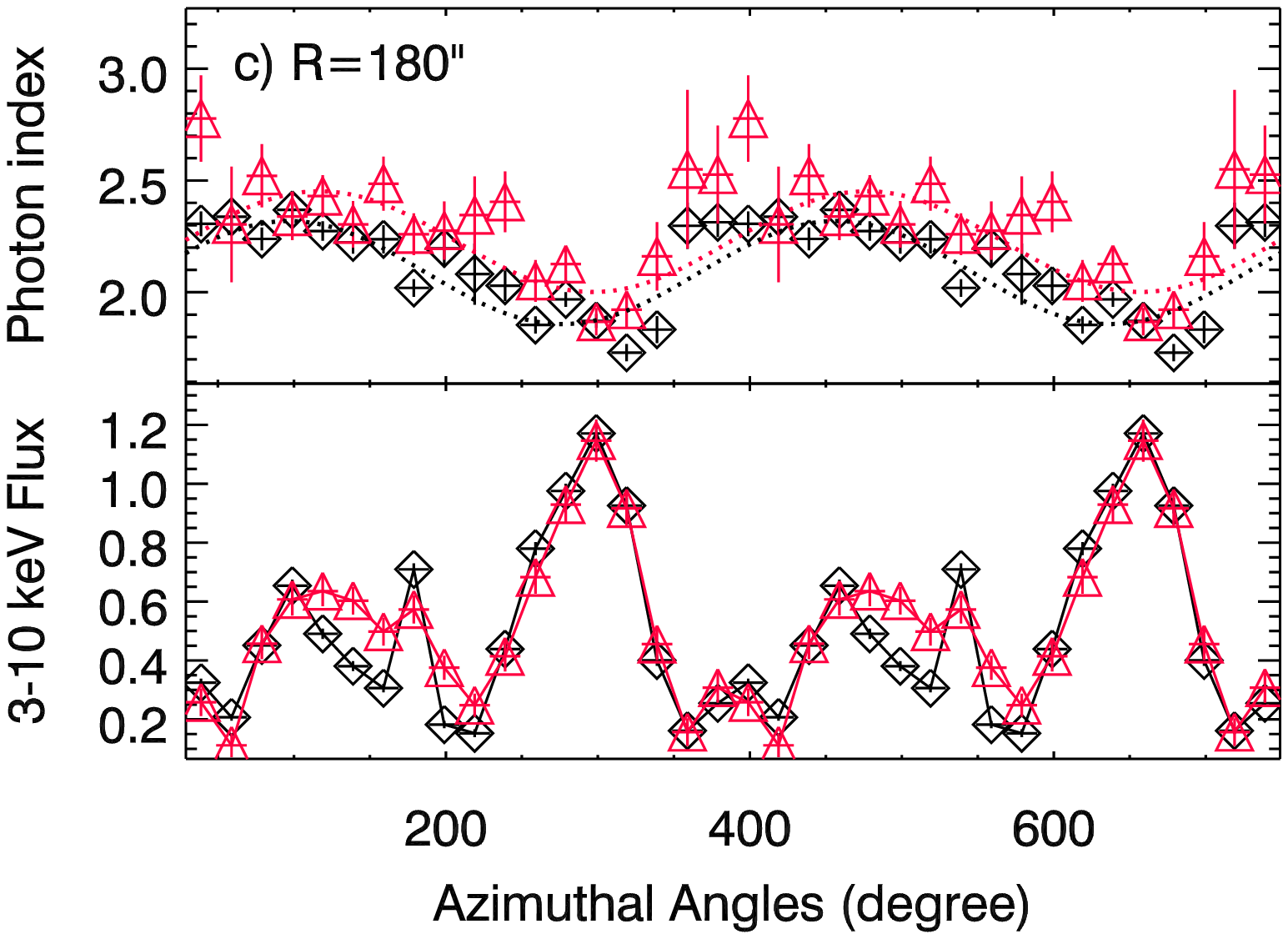} \\
\vspace{2.0 mm}
\hspace{-3.0 mm}
\includegraphics[width=2.45 in]{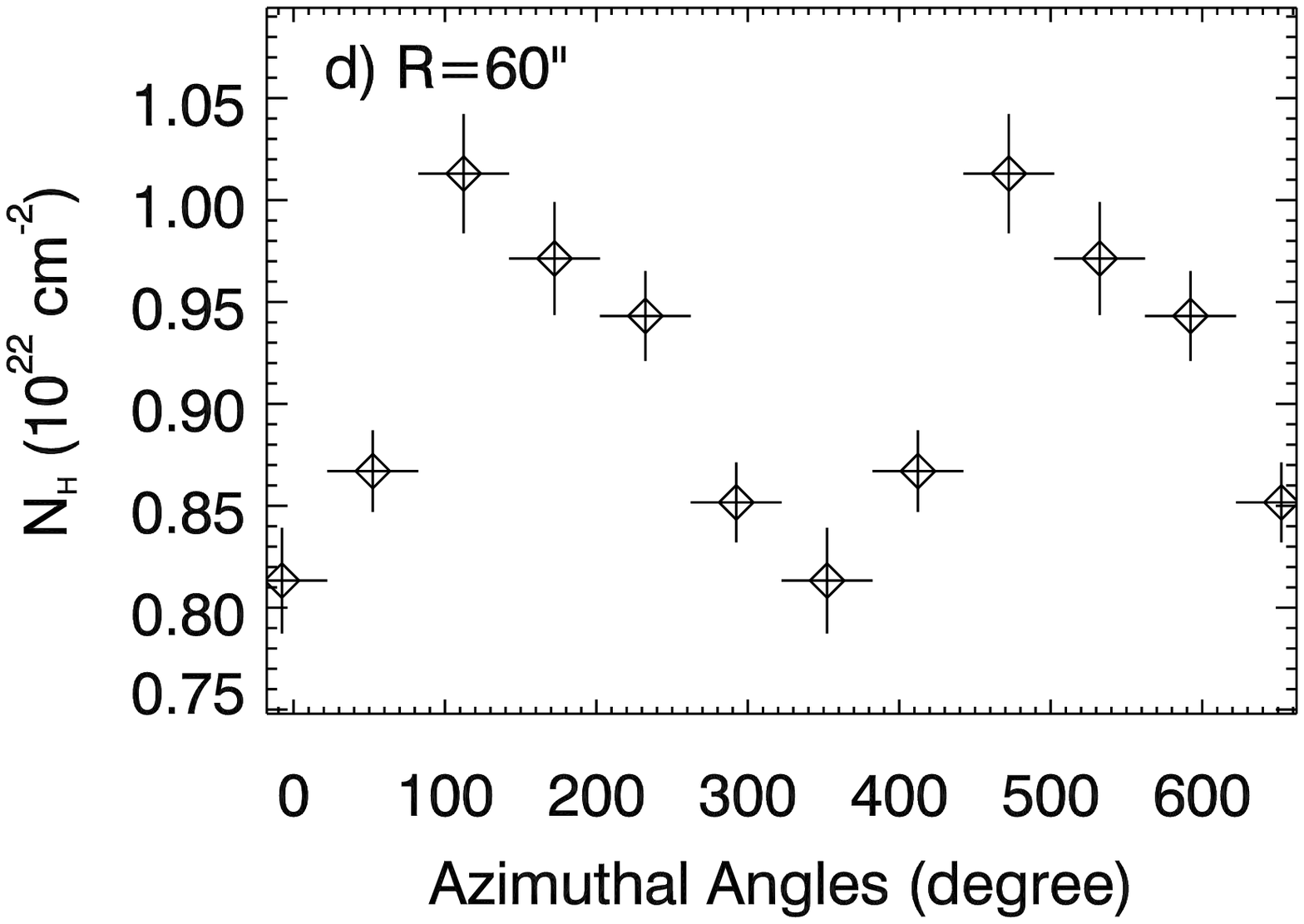} &
\hspace{-8.0 mm}
\includegraphics[width=2.45 in]{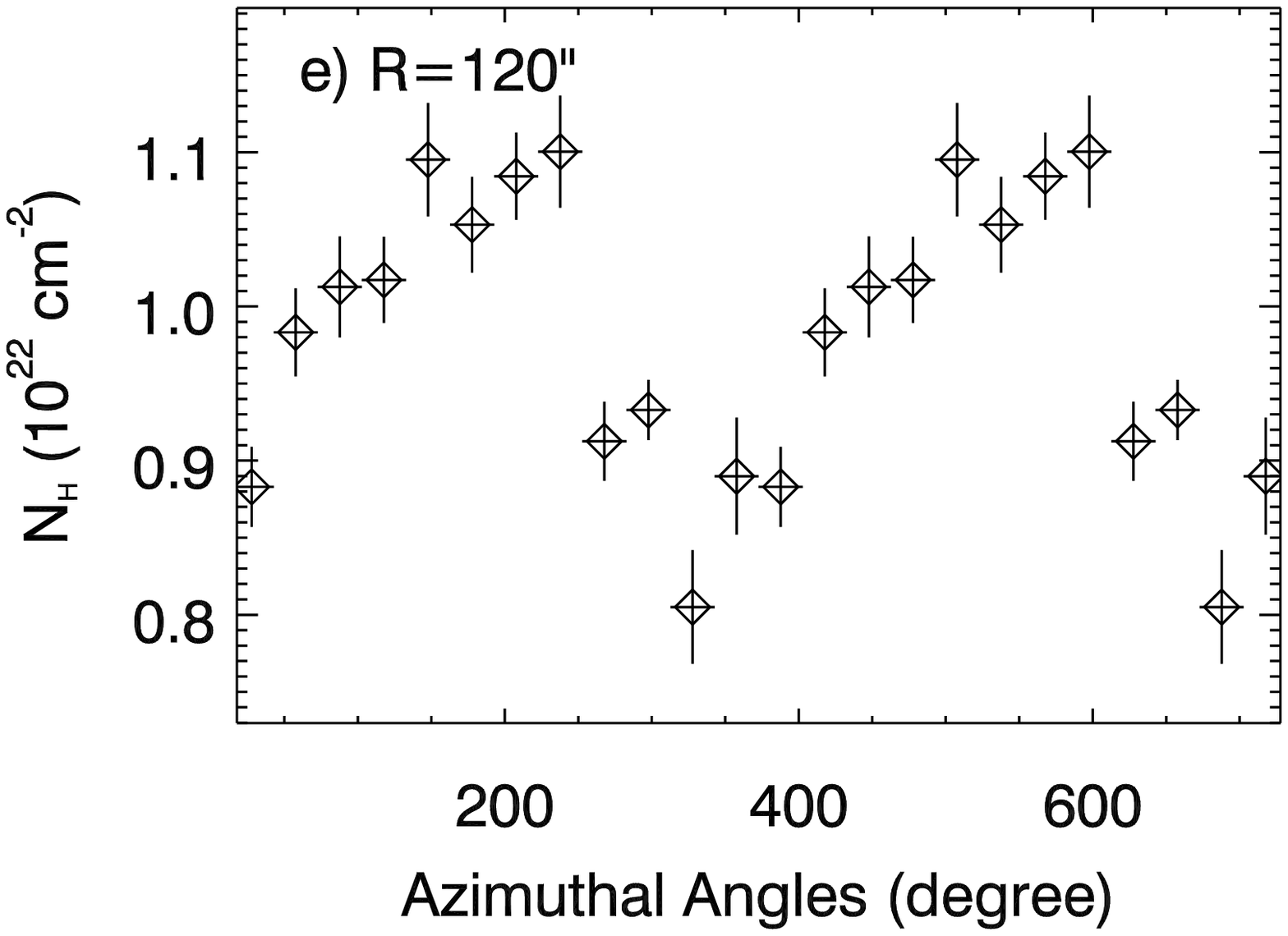} &
\hspace{-8.0 mm}
\includegraphics[width=2.45 in]{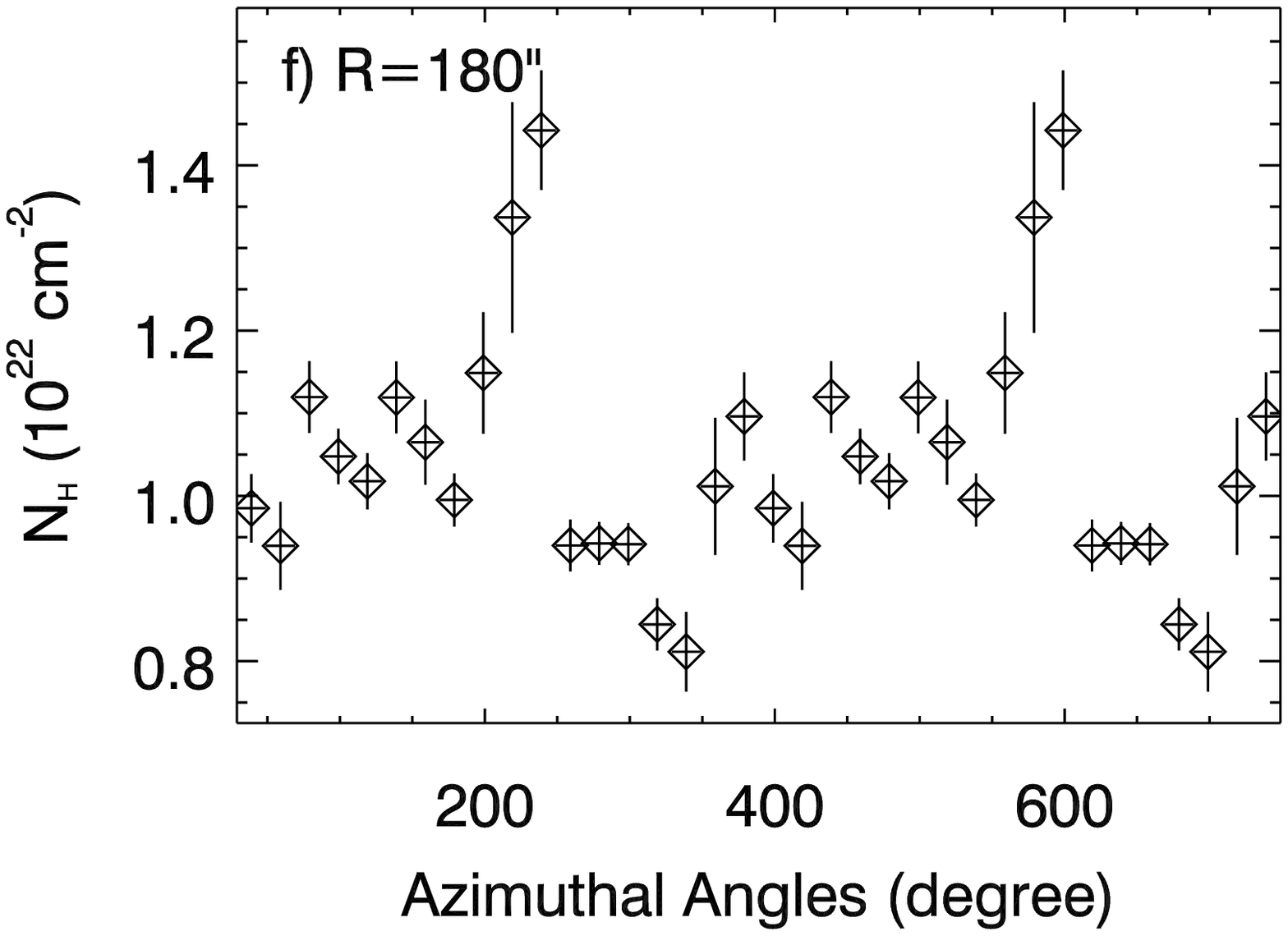} \\
\end{tabular}
\vspace{0.0 mm}
\figcaption{Azimuthal variation of spectral hardness and flux measured with {\em NuSTAR} (triangles)
and {\em Chandra} (diamonds) at $R=60''$ (a), $R=120''$ (b), and $R=180''$ (c) from the pulsar.
Sinusoidal trends for the photon index variation are shown in dotted lines, and solid lines connecting
flux data points are shown for clarity. Flux is in units of $10^{-12}\ \rm erg\ s^{-1}\ cm^{-2}$.
$N_{\rm H}$ values measured with {\em Chandra} for the same regions
are shown in (d)-(f).
\label{fig:AzVar}
\vspace{2.0 mm}
}
\end{figure*}

We compared the {\em NuSTAR} and the {\em Chandra} spectra
in the same energy range below the \AR{possible} break at 6~keV
in order to see if the break is a single sharp break. If so, we expect the {\em NuSTAR} and
the {\em Chandra} spectra to be same regardless of the difference in the effective area shape.
We fit the {\em NuSTAR} and the {\em Chandra} spectra in the common energy band,
defined as 3--$E_{\rm max}$,
where we vary $E_{\rm max}$ from 4.5~keV to 7~keV. In the lowest energy band (3--4.5~keV), the
{\em NuSTAR} and {\em Chandra} results agreed with photon indices of 1.99(7) and 1.98(2).
However, as the upper energy range increased to 5~keV and above, the {\em NuSTAR} results
were significantly softer than that of {\em Chandra}.
For example, the photon indices are 2.03(2) and 1.90(1) in the 3--7~keV band, for {\em NuSTAR} and
{\em Chandra} respectively, inconsistent with each other with 90\% confidence.
This suggests that \AR{the spectral break is likely to be caused by the cross-calibration effect.
However, if the spectral break is real, the observational discrepancy
in the common energy band may imply that}
the broadband spectrum is not sharply broken at 6.3~keV but slowly curves
over a energy range (e.g., 4--7~keV) probably because different regions in the 300$''$ aperture
have different break energies.
In this case, {\em NuSTAR} collects relatively more photons above the break than {\em Chandra} does
since it has rising effective area in that band, yielding a softer spectrum.

\medskip
\subsubsection{Azimuthal Variation of the Spectrum}
In order to see if the PWN spectrum varies azimuthally, we first extracted {\em NuSTAR}
events in 30$''$ radius circles for six, twelve, and eighteen azimuth angles for three radial
distances, 60$''$, 120$''$, and 180$''$ from the pulsar, respectively.
The regions do not overlap. For each region, backgrounds were extracted
from an aperture of $R=45''$ in a source-free region on the same detector chip.
We jointly fit the {\em NuSTAR} spectra for the three observations N1, N2 and N4.
The energy ranges for the fit were 3--20~keV, where the
source events were detected above the background.
We performed these analysis for the same regions with the {\em Chandra} data
in the 0.5--7~keV band.

We show the results in Figure~\ref{fig:AzVar}, where the azimuth angle $\phi$ is
defined from east in a clockwise direction.
A sinusoidal variation of the spectral hardness is clearly
visible for each radial group, and the spectrum is hardest in the jet region ($\phi \sim 300^{\circ}$).
The flux values also peak in the jet regions but do not seem to vary sinusoidally.
Note there is a small discrepancy between the {\em NuSTAR} and {\em Chandra} measurements
in Figure~\ref{fig:AzVar}; the {\em NuSTAR}-measured spectral indices are larger than
those measured with {\em Chandra} in general.
While this may suggest the spectral break we see
in the spectrum of entire nebula (Section~\ref{sec:totspec}),
it could be due to PSF mixing in the {\em NuSTAR} data;
when there is a sharp spatial contrast such as the jet
or image edge, it is convolved with the PSF in the {\em NuSTAR} data.

It appears that the photon index covaries with $N_{\rm H}$ in Figure~\ref{fig:AzVar}. We checked
if there is a correlation between the two quantities using Pearson's product moments, and found
no clear correlation in any radial group.

\medskip
\subsubsection{Spectral Variation in the Northern Nebula and in the Jet}
\label{sec:radspec}
\begin{figure*}
\centering
\hspace{0.0 mm}
\includegraphics[width=0.35\textwidth]{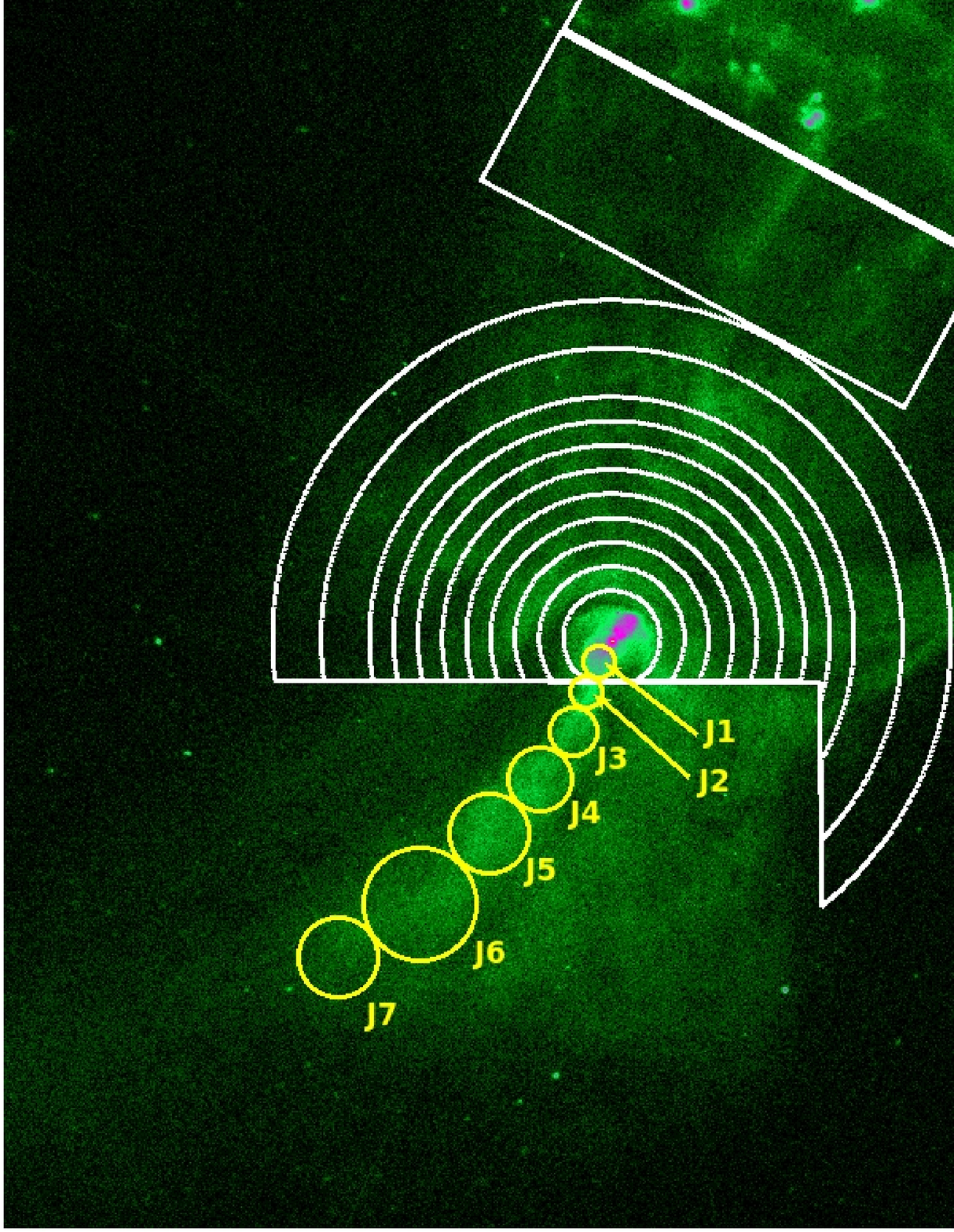}
\hspace{1.0 mm}
{\vbox{\offinterlineskip\halign{#\hskip 4mm&#\cr
  \includegraphics[width=0.30\textwidth]{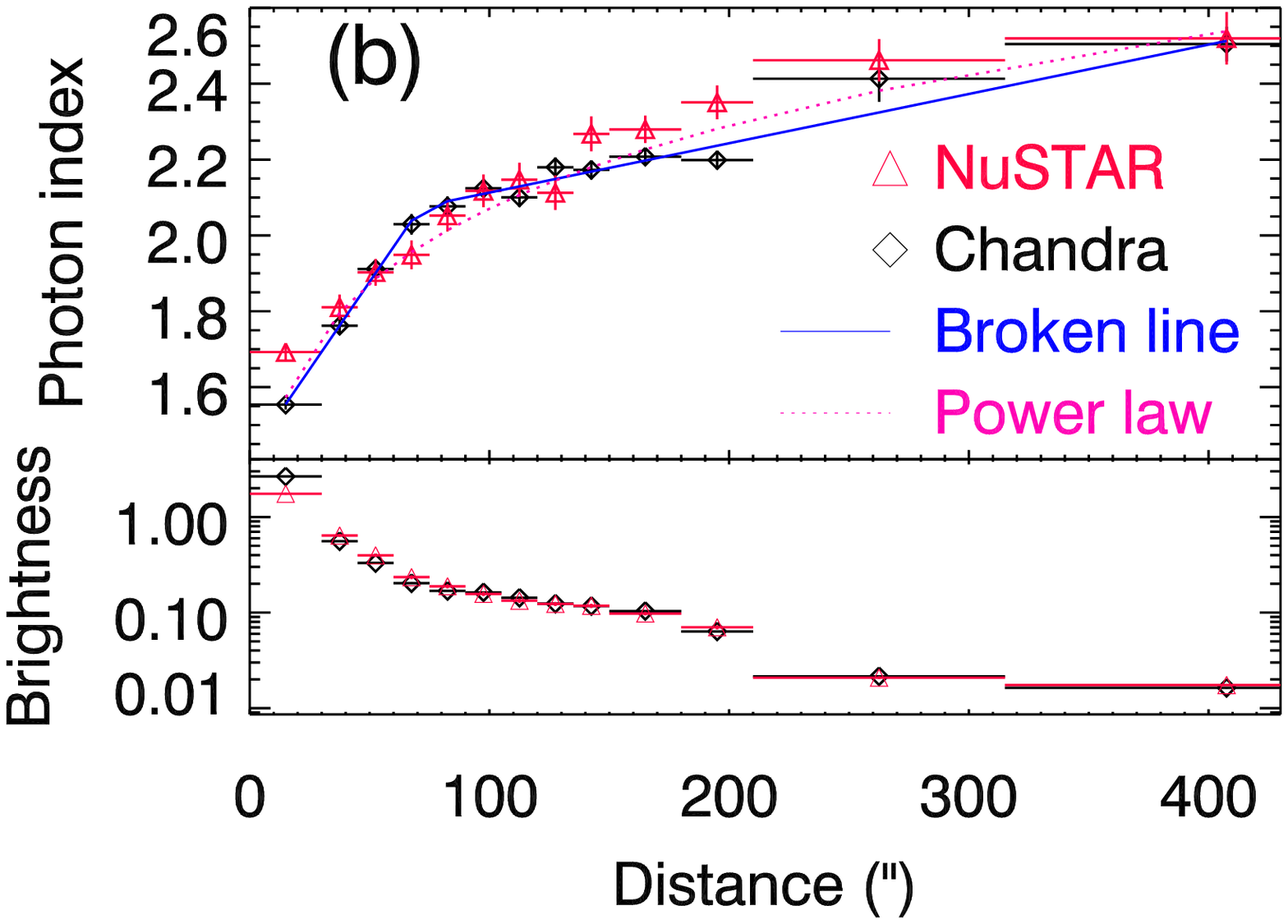}&
  \includegraphics[width=0.30\textwidth]{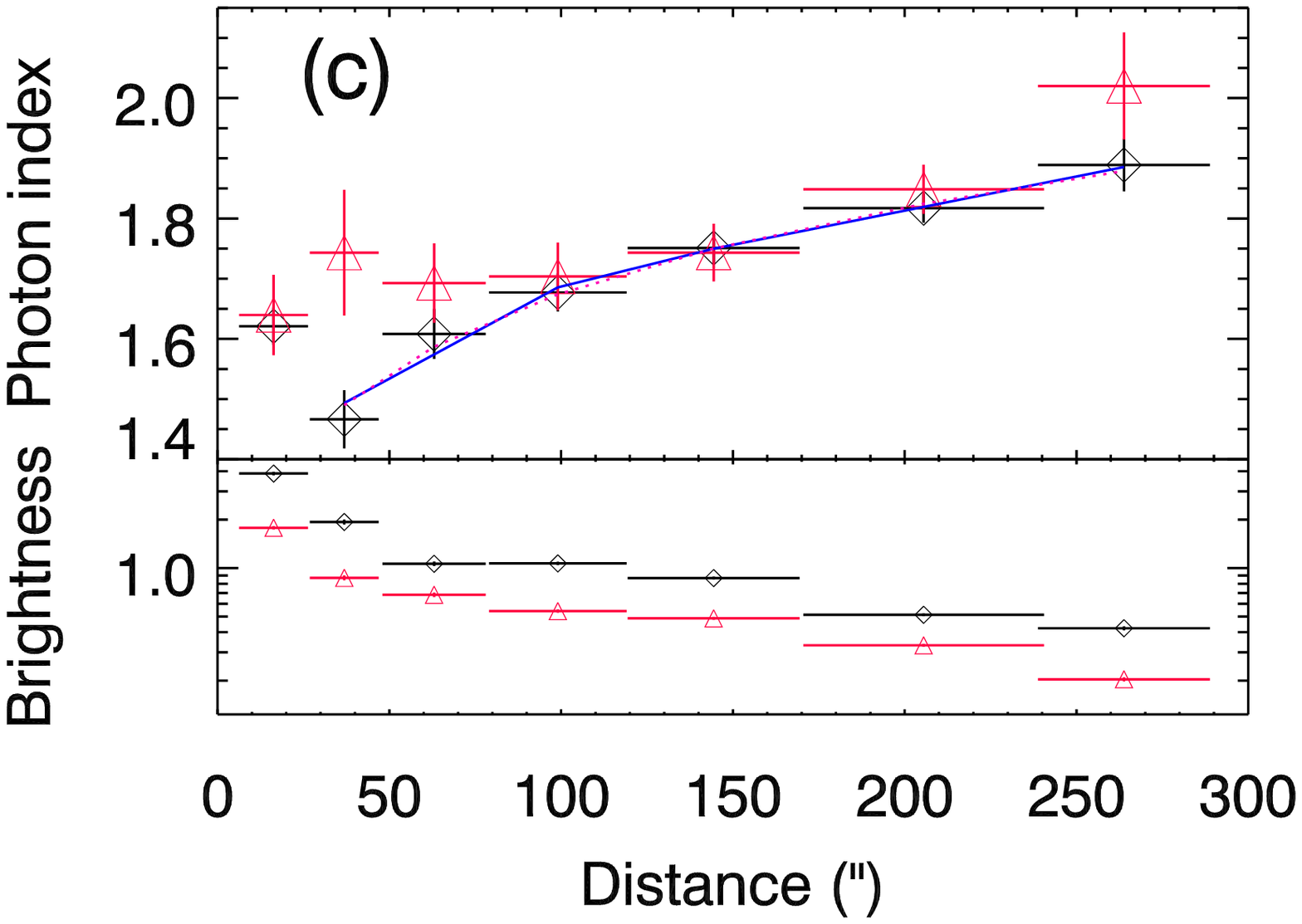}\cr
  \noalign{\vskip 0.0mm}
  \hspace{0.0 mm}
  \includegraphics[width=0.30\textwidth]{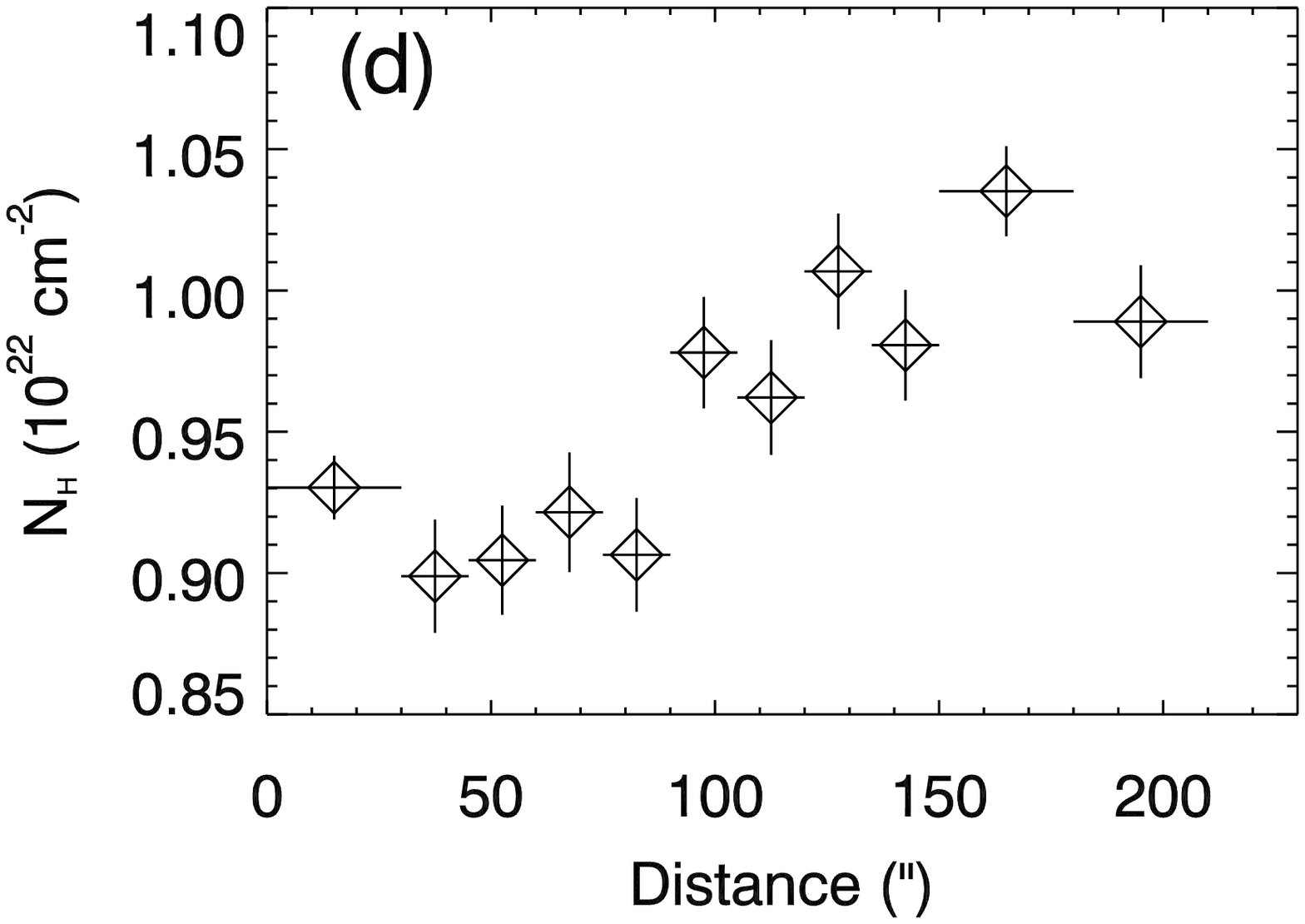}&
  \includegraphics[width=0.30\textwidth]{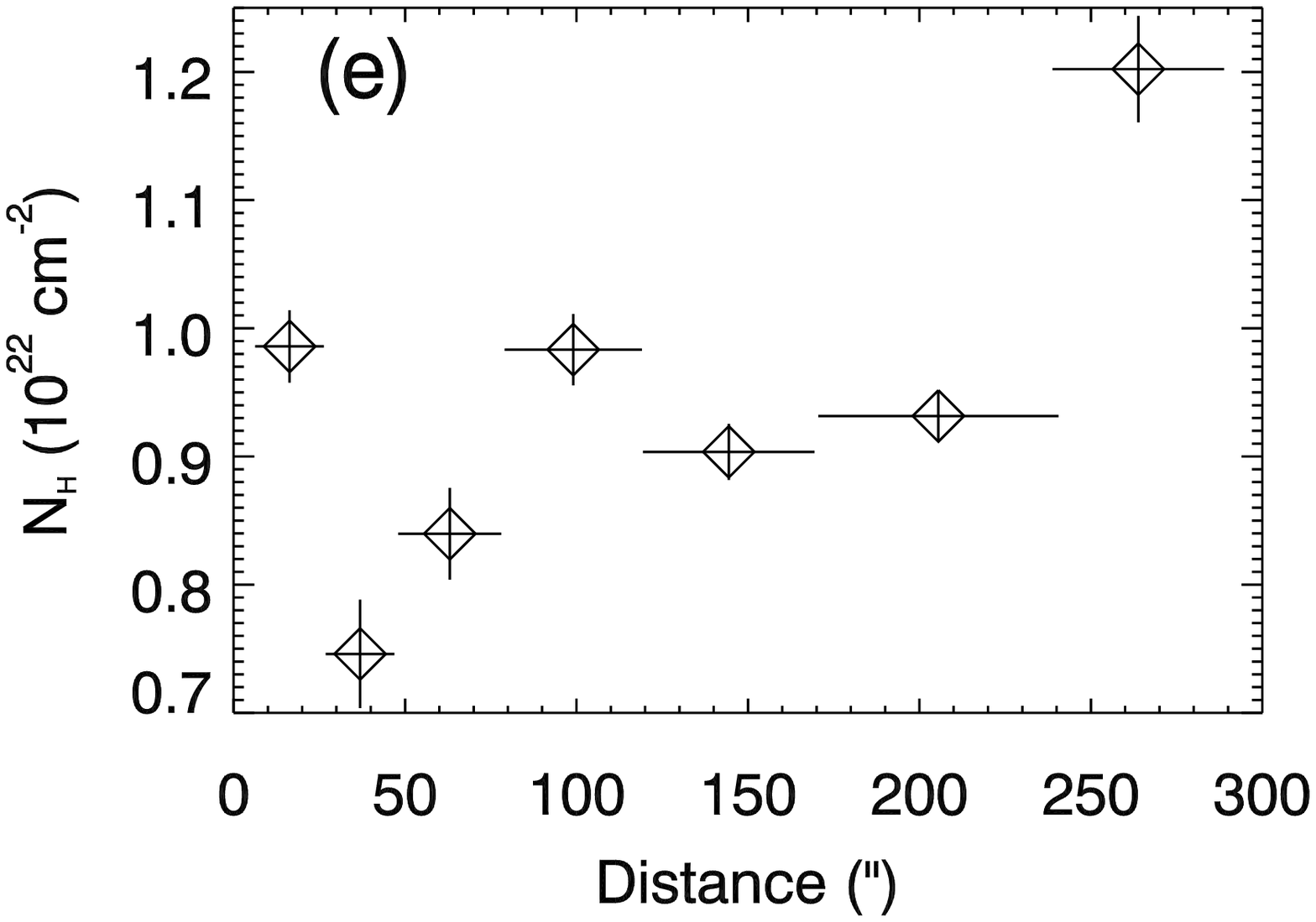}\cr
  }}}
\vspace{2.0 mm}
\figcaption{Selected regions for studying the spectral variation
in the northern nebula (white annuli and boxes) and in the jet (yellow circles) (a),
and spectral variation in the northern and the southern jet directions (b)--(e).
Radial variation of the photon indices and surface brightness in the northern nebula (b) and
in the jet (c) measured with {\em NuSTAR} (triangles) and {\em Chandra} (diamonds). Also shown are
the best-fit broken line (blue solid line) and the power law (magenta dotted line).
Legends are same for (b) and (c).
Brightness is measured in the 3--10~keV band in units of
$10^{-15}\ \rm erg\ s^{-1}\ cm^{-2}\ arcsec^{-2}$.
Radial profiles of $N_{\rm H}$ measured with {\em Chandra} are shown in (d)
for the northern nebula and in (e) for the jet. Note
that the range for the x-axis of (d) is different from that of (b) because $N_{\rm H}$ was not measured for
the last two data points in (b).
\label{fig:RadVar}
}
\end{figure*}

Since we observe significant spectral variation in the azimuthal
direction, we analyzed the northern nebula and the jet separately.
For the northern nebula we used annular regions, ignoring the southern part.
Hence, each region covers the upper $\gapp$180$^{\circ}$ in azimuth angle.
The innermost region was a circle with radius 30$''$ centered at the pulsar, and
annuli with width 30$''$ or 60$''$, and boxes
were used for the outer regions (see Fig.~\ref{fig:RadVar}a white).
We used the off-pulse phase only for {\em NuSTAR}, and ignored the central pulsar
using a circle with radius $5''$ for {\em Chandra}.

We separately fit the {\em NuSTAR} (N1, N2, and N4) and the {\em Chandra} (C1--C5) spectra
of each region with an absorbed power-law model.
Since there is significant thermal contamination from RCW~89 in the {\em Chandra} data at large
distances (see the two upper white box regions in Fig.~\ref{fig:RadVar}a),
we ignore the low energy data below $\sim$3~keV for the two rectangular regions for the {\em Chandra} fits.
We also tried to model the RCW~89 regions
using the {\ttfamily vnei} plus a power law in {\ttfamily XSPEC},
and fit the {\em Chandra} data in the 0.5--7~keV band.
The result for the photon index was sensitive to the remnant model
but broadly agree with what we found by fitting data above 3~keV only \citep[see also][]{ykk+05}.
Our results for the {\em Chandra} data are consistent with,
but more accurate than those obtained by \citet{dga+06} who used
$\sim$60 ks of observations taken from 2000 August to 2003 October.
We present our measurements in Figure~\ref{fig:RadVar}b and d.

While our results show that $N_{\rm H}$ increases with radius,
we note that it is possible to force the $N_{\rm H}$ to be constant
and allow only the photon index to vary.
For example, an $N_{\rm H}$ value of $0.957\pm 0.005\times 10^{22}\ \rm cm^{-2}$ constant over the field
with photon indices of 1.58--2.16 fits the data out to $R=200''$ in Figure~\ref{fig:RadVar} ($\chi^2$/dof=12289/12363),
which implies no radial variation of $N_{\rm H}$ and smaller variation
of the photon index with radius. However, the $N_{\rm H}$ profile given in Figure~\ref{fig:RadVar}d
provides better fit with $\chi^2$/dof=12223/12353 corresponding to F-test probability of $2\times10^{-10}$.
We also verified that the results do not change if we fit the spectra only in the 0.5--6~keV
(below the spectral break, see section~\ref{sec:totspec}) in order to reduce the effect of the complex continuum.
We note that better constraining $\Gamma$ and $N_{\rm H}$ by jointly fitting the {\em NuSTAR} data is
not possible because of the PSF mixing in the {\em NuSTAR} data and the spectral break (Section~\ref{sec:totspec}).

There is a hint of a possible break in the linear slope
of the radial profile of the photon index $\Gamma(R)$ (Fig.~\ref{fig:RadVar}b).
We measured the location of the break in the northern nebula using a broken line fit.
We first fit the {\em Chandra} measured
photon index profile, and found that the break occurs at $R_{\rm break}=71 \pm 3''$.
We note that using a constant $N_{\rm H}$ over the field changes $\Gamma$ only slightly and
gives a consistent result ($R_{\rm break}=68 \pm 2''$).
The {\em NuSTAR} profile gives a larger $R_{\rm break}=150\pm10''$ because of the large photon index at
smaller radii which might be biased by mixing from outer regions.
We also find that a single power-law model $\Gamma(R)=\Gamma_0 R^\eta$ with
$\eta=0.149\pm0.003$ broadly agrees with the data (see Fig.~\ref{fig:RadVar}b).

Since spectral softening is expected in the jet direction as well,
we measured the spectral variation along the southern jet. To do this,
we extracted source spectra using non-overlapping circular apertures
with radii 10$''$, 10$''$, 15$''$, 20$''$, 25$''$, 35$''$, and 25$''$ along the jet
(see yellow circles in Fig.~\ref{fig:RadVar}a), which we refer to as regions J1--J7.
Note that the center of J1 is $R\sim15''$ from the pulsar, and all the {\em NuSTAR} observations (N1--N4)
were used for this analysis. We fit the spectra in each region with an absorbed power law,
and measured the photon index and flux. The results are presented in Figures~\ref{fig:RadVar}c and e.

The spectral indices of the J2 region measured with {\em Chandra} and {\em NuSTAR} are very different,
which may suggest that there is a strong spectral break. However,
we note that measuring the spectral parameters with {\em NuSTAR} was difficult for regions with
sharp spectral changes because the {\em NuSTAR} PSF changes from a circular shape to an elliptical shape with 
off-axis angle \citep{amw+14}, and thus regions with different off-axis angles
have different degrees of azimuthal mixing. The four {\em NuSTAR} observations had
different pointings and thus different off-axis and azimuthal angles. In particular, in the regions J1--J2 where
we use small apertures and the source spectrum strongly varies, spatial mixing has significant impact
on the {\em NuSTAR} results. Therefore, the discrepancies in the spectral index between
the {\em NuSTAR} and {\em Chandra} measurements, and even between the {\em NuSTAR} observations
are expected. The mixing was not a concern in the analysis of the northern nebula in
which spectral variation is not severe.

We also measured the location of the break in the radial profile of the photon index in the
jet direction using the {\em Chandra} measurement in Figure~\ref{fig:RadVar}c.
Here we ignored the first data point for the reason described below.
A fit to a broken line gave a break location of $R_{\rm break}=110\pm30''$.
A single power-law model also fits the data with a power index $\eta=0.12\pm 0.01$ (Fig.~\ref{fig:RadVar}c).

We note that the first {\em Chandra} data point, corresponding to region J1,
shows a very soft spectrum compared to the next one in J2, unexpected in
synchrotron cooling models \citep[e.g.,][]{r03, tc12}. The steady-state solutions may not
be applicable to the inner region within $\sim$1$'$ of the pulsar for this source,
as \citet{dga+06} found strong variability in the brightness and morphology in that region.

\begin{figure}
\hspace{0.0 mm}
\includegraphics[width=3.4 in]{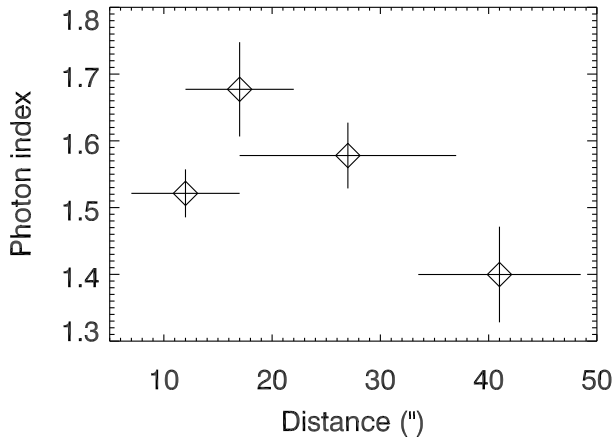}
\figcaption{The spectral indices in the central regions, corresponding
to the innermost two data points of Fig~\ref{fig:RadVar} ($R<50''$), with a higher spatial resolution.
\label{fig:innerjet}
}
\end{figure}

We tried to see if the spectral hardness in J1 varied over time. We first jointly fit
the {\em Chandra} spectra of the region taken from the five observations C1--C5
with a common $N_{\rm H}$, but separate
photon index and cross normalization for each observation, and found that the photon indices
are all within the 1$\sigma$ uncertainty of the value in Figure~\ref{fig:RadVar}. We carried out
the same analysis for the J2 region, and found that the spectrum of one observation
(C1, Obs. ID 754) was slightly softer than the others ($\Gamma=1.62\pm0.12$) but not significantly.
It is probably because the region in this observation fell on the detector chip gap. Therefore,
we conclude that the spectral index did not change significantly over time in this region.

Since spectral hardness covaries with $N_{\rm H}$, the spectral difference
between the J1 and J2 regions may be less significant if we consider the covariance.
In order to investigate the effect of covariance, we ignored Obs. ID 754
because the J2 region in this observation was on the chip gap. We
then jointly fit the spectra of each region, varied both $N_{\rm H}$ and $\Gamma$ using
the {\ttfamily steppar} tool in {\ttfamily XSPEC} and found that the 99\% contours do not overlap,
which suggests that the difference is significant with the covariance as well,
and the spectrum of the J1 region is significantly softer than that of the J2 region.
If we take the best-fit values, the $N_{\rm H}$ variations of $\sim 3\times10^{21}\ \rm cm^{-2}$
imply extremely high densities of $n\sim2000\ \rm cm^{-3}$ for an assumed
line-of-sight distance of 0.5 pc (similar to the transverse distance
for the assumed distance to the source of 5.2~kpc)
in the regions with high $N_{\rm H}$.

\begin{figure*}[htb]
\hspace{-8.0 mm}
\begin{tabular}{ccc}
\includegraphics[width=3.05 in]{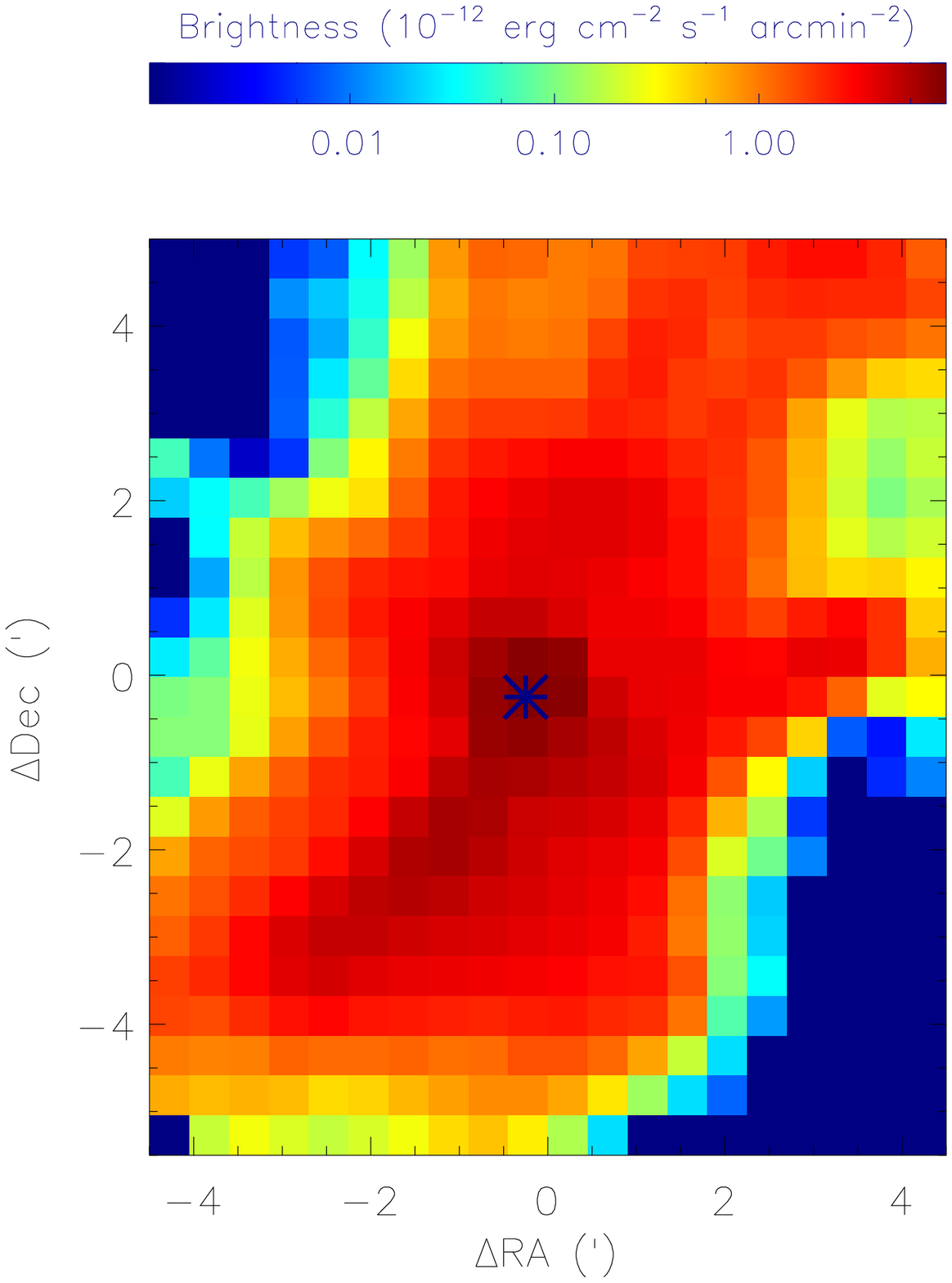} &
\hspace{-22.0 mm}
\includegraphics[width=3.05 in]{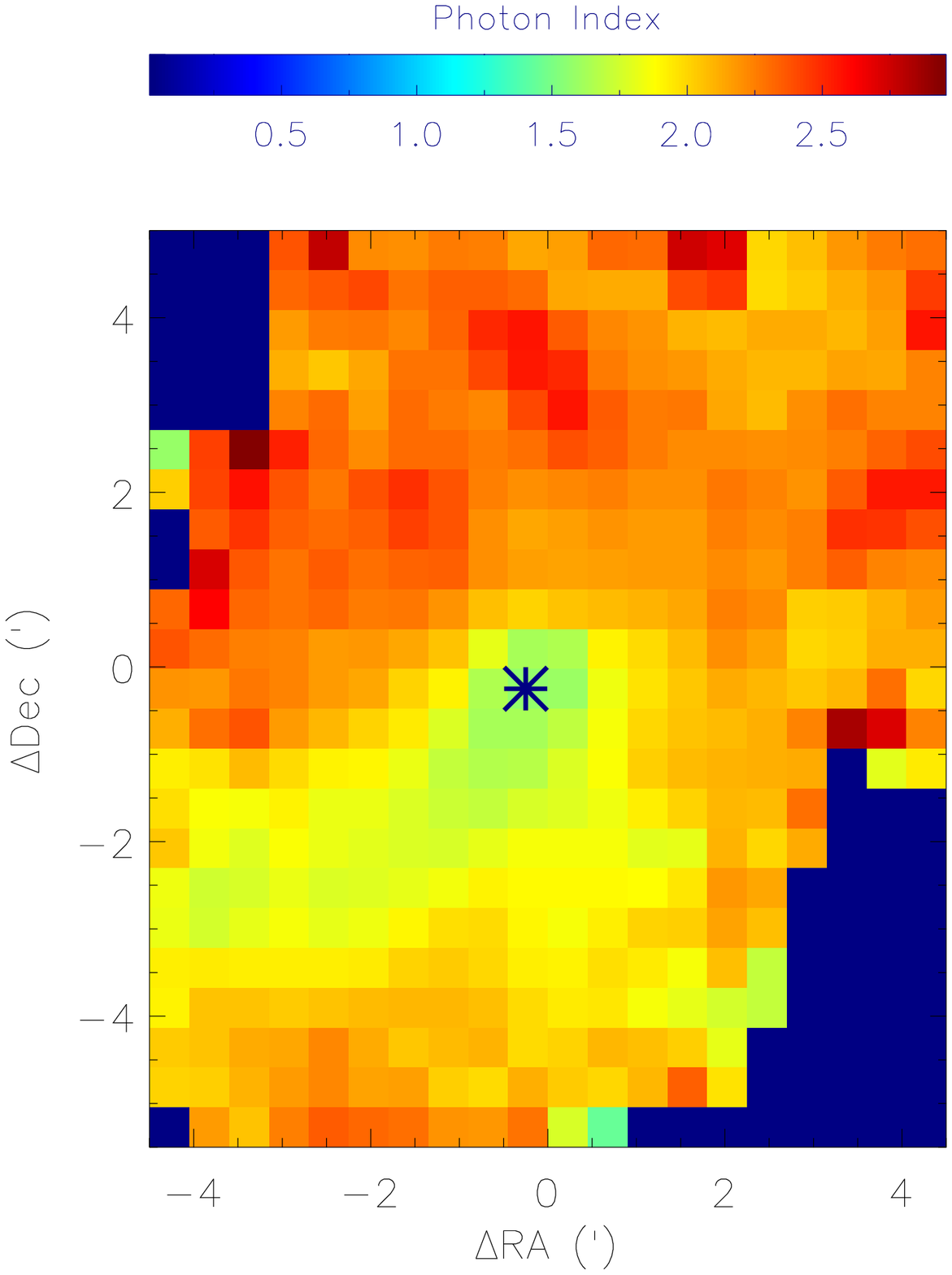} &
\hspace{-22.0 mm}
\includegraphics[width=3.05 in]{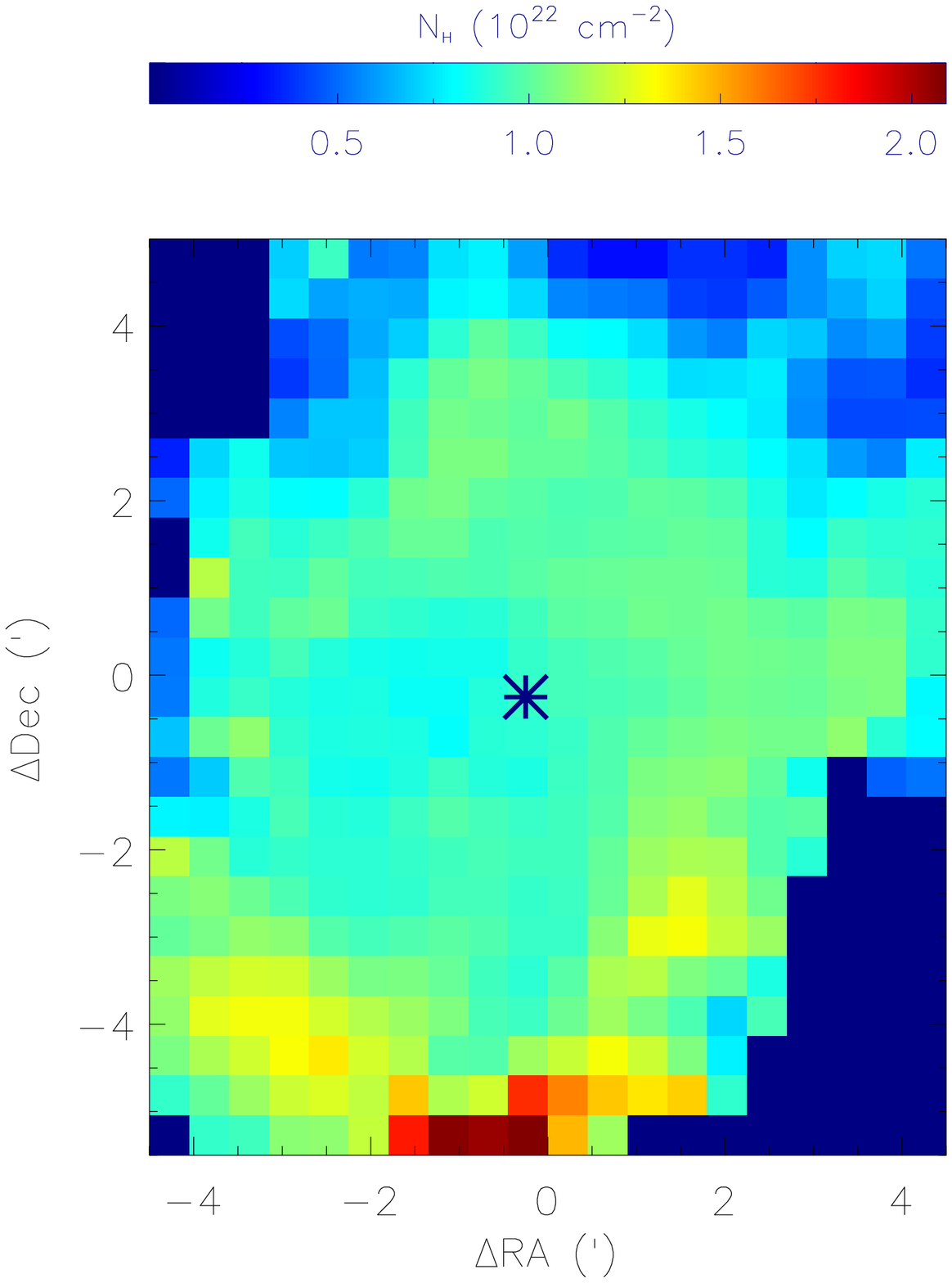} \\
\includegraphics[width=3.05 in]{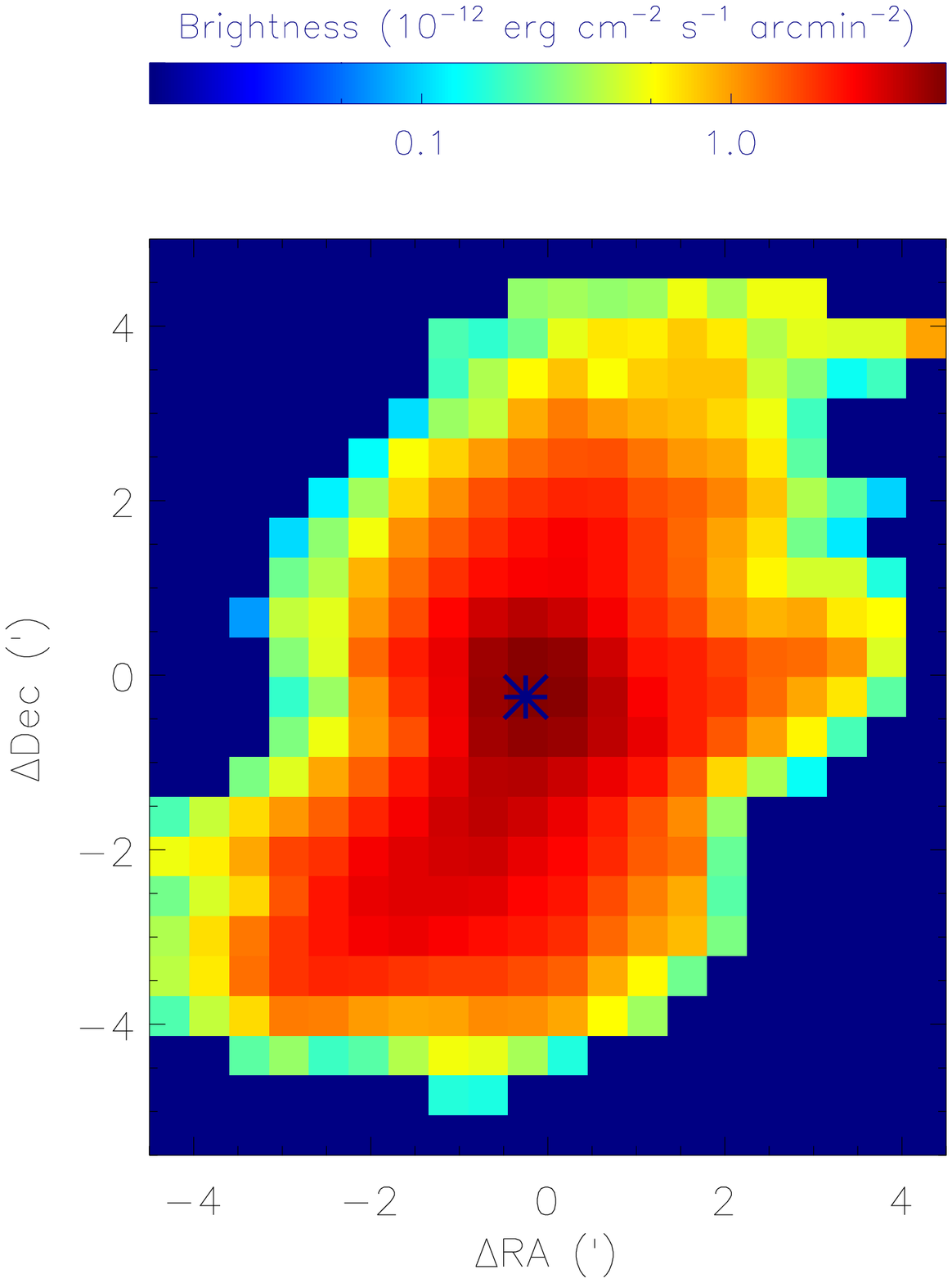} &
\hspace{-22.0 mm}
\includegraphics[width=3.05 in]{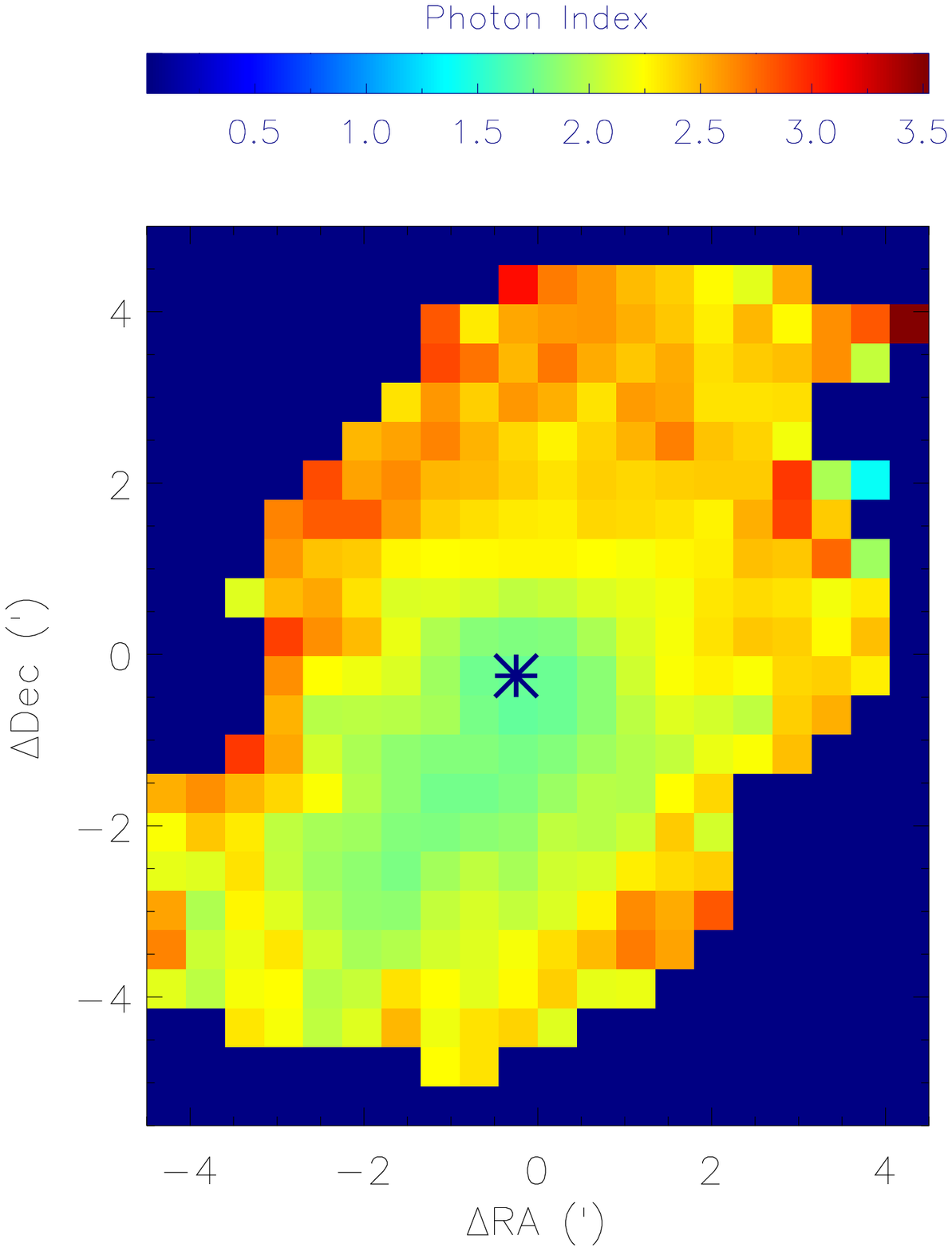} &
\hspace{-22.0 mm}
\includegraphics[width=3.05 in]{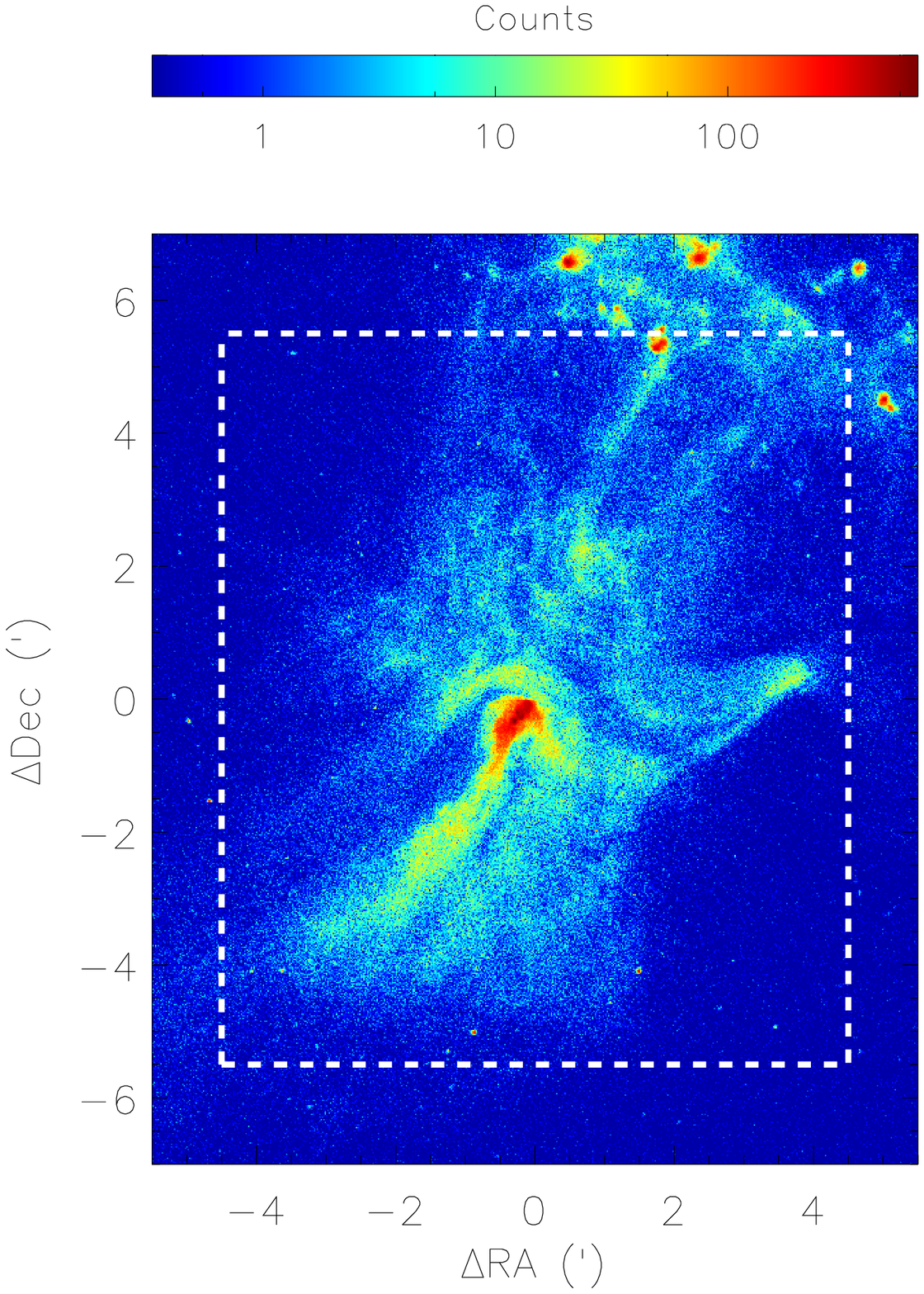} \\
\end{tabular}
\figcaption{{\it Top:} 2-D maps of 3--10~keV
brightness in units of $10^{-12}\ \rm erg\ cm^{-2}\ s^{-1} arcmin^{-2}$,
power-law photon index, and $N_{\rm H}$ (from left to right) measured with {\em Chandra}
by fitting the spectra in the 0.5--7~keV band.
{\it Bottom left and middle:} Brightness and photon index measured with {\em NuSTAR}
by fitting the spectra in the 3--20~keV band.
{\it Bottom right:} {\em Chandra} counts map and the field for the 2-D maps (white dashed box).
The location of the pulsar, PSR~B1509$-$58, is noted with a star except for in the bottom right plot.
We arbitrarily assigned zero values to the spectral parameters in regions where the parameters
are unconstrained due to paucity of source counts (dark blue regions).
\label{fig:2Dmap}
}
\end{figure*}

We further spatially resolved the J1--J2 regions using overlapping circular regions with radius 5$''$.
We fit the spectrum of each region with a power-law model, and found spectral softening in the
innermost regions ($R\lapp 20''$). We show the photon indices in Figure~\ref{fig:innerjet}.

Note that $N_{\rm H}$ increases with radius in the northern nebula. In the jet direction, we
used finer spatial resolution, and see a more complicated change; there is a dip at $R$=30--70$''$.
At large distances, we find that $N_{\rm H}$ is large. This structure is visible in the 2-D $N_{\rm H}$
map as well (see Fig.~\ref{fig:2Dmap}). Note also that the power-law index ($\eta$)
of the photon index profile is larger in the northern nebula than in the jet,
that is, the spectral steepening is more rapid,
which was also implied by the imaging analysis above (e.g., Fig.~\ref{fig:comp}).

\medskip
\subsubsection{2-D Maps of the Spectral Parameters}
\label{sec:2dmap}
We produced 2-D maps of the spectral parameters for the $\sim$9$'\times$11$'$ field
containing the PWN.
We used a 1$'\times$1$'$ square region, sliding it over the field with a step of 0.5$'$.
Thus, two adjacent regions overlap by 50\%.
Backgrounds were extracted from far outer regions. In order to minimize the pulsar contamination,
we excluded a circular region with radius 5$''$ for the {\em Chandra} data,
and used the pulse phase 0.7--1.0 only for the {\em NuSTAR} data.

After extracting 0.5--7~keV spectra in each region for the five {\em Chandra} observations, 
we jointly fit the spectra with a common absorbed power-law model having different
cross normalization factors between observations and
allowing all the parameters to vary throughout.
The same procedure was applied to the {\em NuSTAR} data in the 3--20~keV band with $N_{\rm H}$ frozen to
the {\em Chandra}-measured value in each region.
After producing the 2-D maps, we select regions with positive flux with 3$\sigma$ confidence
and show the results in Figure~\ref{fig:2Dmap}.
The average (median) of the 1$\sigma$ uncertainties for the parameters
obtained with the {\em Chandra} data were
$1.8\times 10^{-14}\ \rm erg\ cm^{-2}\ s^{-1}$ ($1.6\times 10^{-14}\ \rm erg\ cm^{-2}\ s^{-1}$),
0.07 (0.05), and $5.6\times 10^{20}\ \rm cm^{-2}$ ($3.6\times 10^{20}\ \rm cm^{-2}$),
for flux, photon index, and $N_{\rm H}$, respectively.
For {\em NuSTAR}, the uncertainties were
$6.0\times 10^{-14}\ \rm erg\ cm^{-2}\ s^{-1}$ ($5.8\times 10^{-14}\ \rm erg\ cm^{-2}\ s^{-1}$)
and 0.15 (0.12) for flux and photon index, respectively.
The flux map shows the structures seen in the count
map (Fig.~\ref{fig:image}). We also note that the photon index map shows the same structure seen
in the radial and the azimuthal profiles (Figs.~\ref{fig:AzVar} and \ref{fig:RadVar});
the photon index increases radially outwards.

We find an interesting shell-like structure in the $N_{\rm H}$ map
(top right panel of Fig.~\ref{fig:2Dmap}).
Since it is possible that the structure is produced by a correlation
between $\Gamma$ and $N_{\rm H}$, we calculated the correlation coefficient
between pairs of parameters using Pearson's product moments. The coefficients were transformed
into Fisher coefficients to calculate the significance. In this study, we found that
the correlation between flux and photon index is $-0.58$, and the significance is 12$\sigma$,
implying the correlation is statistically significant.
No correlation was found between $N_{\rm H}$ and $\Gamma$ or any other combination of parameters.

We further tried to fit the {\em Chandra} data with a single
$N_{\rm H}$ value of $0.95\times 10^{22}\ \rm cm^{-2}$ for the entire nebula,
and found that the fits became worse, having average $\chi^2$ per average dof of 555/547 compared
to the value of 538/546 for the fit with variable $N_{\rm H}$.
The single $N_{\rm H}$ fit turned the large $N_{\rm H}$ regions into spectrally hard regions
which are not visible in the {\em NuSTAR} map.
However, we note that quantitative comparison with the {\em NuSTAR} map is difficult
unless we know the details of the \AR{broadband spectrum in each region}.

\medskip
\section{Discussion}
\label{sec:disc}
We have presented the first hard X-ray images of MSH~15$-$5{\sl2} above $\sim$8~keV.
The broadband images show the synchrotron burn-off effect which is asymmetric in space
(see Figs.~\ref{fig:image} and \ref{fig:comp}).
We \AR{find a possible spectral break at $E_{\rm break}=6\ \rm keV$}
in the spectrum of entire PWN.
From the spatially resolved spectral analysis, we found that the spectral
index varies sinusoidally in the azimuth direction
from $\sim$2 in the north to 1.6 in the south at a distance of $60''$ (1.5 pc) from the pulsar
and monotonically increases with distance from 1.6 to 2.5 (Fig.~\ref{fig:RadVar}).
These trends were observed with both {\em NuSTAR} and {\em Chandra}.
We found that spectral hardness turns over at $R=35''$ and decreases more slowly beyond $R\sim 60''$
along the jet (Fig.~\ref{fig:RadVar}), and showed that there is a previously unrecognized
shell-like structure of radius $\sim3'$ in the $N_{\rm H}$ map (Fig.~\ref{fig:2Dmap}).

\medskip
\subsection{Image}
\label{sed:imgdisc}
The {\em NuSTAR} images in the hard band ($\gapp$7~keV, Figs.~\ref{fig:image}e--h)
show that the source shrinks with energy in the 2-D projection on the sky,
which is attributed to the synchrotron burn-off effect. While
the effect has been shown for this source
in a previous study of azimuthally integrated spectra \citep[][]{sbd+10},
this is the first time it is shown with a 2-D imaging analysis
over a broad X-ray band. In particular, we found
that the burn-off effect is stronger in the northern direction than in the jet direction (Fig.~\ref{fig:comp}).
Although it is very difficult to build a full 2-D model that can reproduce the measured images,
matching overall morphological changes in energy may give us important clues to understand
particle outflow in the PWN.

The change of the size with energy can be used for inferring properties of the particle
flow in the PWN using theoretical models \citep[][]{kc84, r09}.
For example, the results of \citet{r09} can be rewritten in a form appropriate
for sources whose spectral break $\Delta$ between radio and X-ray power laws,
and size index $m$ can both be measured:
$${\Delta \over (-m)} = \left(1 \over \epsilon \right)
  \left(1 + 2\epsilon + (3 + 2\alpha_r) m_\rho/3 + (1 + \alpha_r) m_b\right),$$
where $m_x$ is the index of an assumed power-law function for
a quantity $x\propto R^{m_x}$ ($x=\rho$ or $b$ for mass density and magnetic-field strength,
respectively),
$\alpha_r$ is the energy index of the radio spectrum, and $\epsilon$ is a confinement
parameter \citep[e.g., $\epsilon=1$ for conical jets,][]{r09}.
For the values we derive for MSH 15$-$5{\sl 2} of $\alpha_r = 0.2$, $\Delta = 0.86,$
and $m = -0.2$ (Section~\ref{sec:imageana}), the formula gives
$$1 + 1.1 m_\rho + 1.2 m_b = 2.3 \epsilon.$$
This condition requires either that the mass in the flow is not
constant \citep[for instance, due to mass evaporated from thermal gas
filaments joining the outflow;][]{l03}, or that magnetic flux
is not conserved (for instance, due to turbulent amplification or
reconnection): either $m_b$ or $m_\rho$, or both, must be positive.
Various combinations of gradients can reproduce our results.  For instance,
a conical flow $\epsilon = 1$ requires $1.1 m_\rho + 1.2 m_b = 1.3$,
approximately met if density is constant and magnetic field rises as
radius -- or vice versa.  A confined flow with $\epsilon = 0.44$ (roughly
parabolic) would have $m_\rho \cong -m_b$, which could be satisfied
with both constant density and constant magnetic field, or with one dropping
as fast as the other increases.

\medskip
\subsection{Spectra of the Entire PWN}
\label{sec:specdisc}
We find that the integrated spectrum of the entire nebula
measured with {\em NuSTAR} is a simple power law with
photon index 2.06 (Table~\ref{ta:spec}).
This is similar to values previously reported based on {\em BeppoSAX} and
{\em INTEGRAL} data \citep[$\Gamma=2.08\pm 0.01$, $2.12\pm0.05$;][]{mcm+01, fhr+06}.
The {\em Chandra}-measured parameters for a single power-law model above 2~keV
imply a harder spectrum than that measured with {\em NuSTAR} (see Table~\ref{ta:spec}),
which is also seen in the spectra extracted for
other inner regions (e.g., Figs.~\ref{fig:AzVar} and \ref{fig:RadVar}).
We note that our {\em Chandra} spectral fit results are consistent with that measured previously
with {\em BeppoSAX} \citep[$\Gamma=1.90\pm0.02$ in the 1.6--10~keV band for a 4$'$ aperture,][]{mcm+01}.

Since the large aperture include many subregions with different spectral parameters,
we expect to see a harder spectrum at high energies if the spectra of the subregions are
simple power laws in the 0.5--20~keV.
However, the 3--20~keV {\em NuSTAR} spectra are much softer than the {\em Chandra} spectra, which
is the opposite to what is expected from a sum of simple power-law spectra.
We find that contamination of the pulsar and the backgrounds can explain only 
$\sim0.01$ of the photon index discrepancy of the {\em NuSTAR} and {\em Chandra} measurements,
while the measured difference is 0.15.
The correction for the pulsar, the RCW~89 contamination and the background variation
in the data seem not to explain the discrepancy between the {\em NuSTAR} and the {\em Chandra} results.

\AR{We find that the discrepancy between the {\em NuSTAR} and the {\em Chandra} measurements is likely
to be caused, at least in part, by imperfect cross calibration of the instruments.}
\AR{However, if the break is real,} it implies a break in the energy distribution
of the shock accelerated electrons at $\sim$200~TeV (having the peak synchrotron power at 6~keV)
for a magnetic field strength of 10~$\mu$G, assuming synchrotron emission (e.g., Equation 5 in TC12).

We note that a spectral break in the X-ray band has recently been reported for G21.5$-$0.9
\citep[][]{nhr+14} and the Crab Nebula \citep{mrh+14} and may be common to other
young PWNe as well. If true, this has important implications on the particle acceleration mechanism
in PWNe. Sensitive broadband X-ray observations of other PWNe will be helpful.

\medskip
\subsection{Spectral Variation}
\label{sed:specvardisc}
Using the broadband X-ray data obtained with {\em NuSTAR} and {\em Chandra},
we find an azimuthal variation of the spectral index. For the PWN~3C~58, \citet{shv+04} suggested
a possible azimuthal variation of the spectral hardness based on a scenario where
current flows out from the pulsar's pole and returns in the equator \citep[][]{b02}.
However, \citet{shv+04} did not find an obvious azimuthal variation
in the PWN~3C~58 within $R\sim2$ pc for a distance 3.2 kpc
we find for MSH~15$-$5{\sl2} at $R$=1.5--4.5 pc.
Furthermore, we find the azimuthal spectral variation in MSH~15$-$5{\sl2}
is likely sinusoidal and different from that of the flux.
This azimuthal spectral variation of the emission
may hint at a large scale current flow, however, it could also be due to azimuthal
diffusion of jet particles in MSH~15$-$5{\sl2}.

A radial change of the spectral index of MSH~15$-$5{\sl2} was reported by \citet{sbd+10}.
While they integrated the spectrum over the full azimuthal angle,
we measured the profiles for the northern and
the jet directions separately because the two regions are different.
We found that significant softening with radius is seen in both directions, more significantly
in the northern region. Interestingly, the radial profile of the photon index (rate of spectral steepening)
flattens with radius as is also seen in 3C~58 \citep[][]{shv+04}. 

An outflow model considering both diffusion and advection was
developed by \citet{tc12} (TC12 hereafter), where they calculated the change of the spectral
index with distance from the central pulsar with an assumed electron injection spectrum and diffusion
coefficient, and were able to reproduce the radial variation of the spectral index
for three compact PWNe, the Crab nebula, G21.5$-$0.9, and 3C~58.
The model has been applied to PWNe where the particle
escape times (the times for particles to diffuse a distance $R$ in the Bohm limit),
are longer than their ages (Equation 2 in TC12):
$$t_{\rm esc}\approx 16,000\left(\frac{R_{\rm PWN}}{2\ \rm pc}\right)^2\left(\frac{E_e}{100\ \rm TeV}\right)^{-1}
\left(\frac{B}{100\ \mu\rm G}\right)\ \rm yr,$$
where $R_{\rm PWN}$ is the radius of the PWN, $E_e$ is the energy of synchrotron emitting particles, and $B$ is
the magnetic-field strength in the PWN.
Using the size $R\sim10\ \rm pc$, $E_e=$ 100--600 TeV \citep[][]{fhr+06, nky+08}, and an
estimation of the magnetic-field strength of 8--17~$\mu\rm G$ \citep[][]{gak+02,aaab+05},
we find that $t_{\rm esc}$ is 5000--7000 yr, greater than the spin-down
estimated age of $\tau_c=1700$ yr. Since the photon spectrum we
are using in this work corresponds to a smaller $E_e$, $t_{\rm esc}$
can be larger than the above estimation. However, we note that there have been suggestions
that the true age of the PWN is $\gapp 6000$ yr based on the association with RCW~89,
larger than the spin-down estimation \citep[e.g.,][]{shm+83}. If so, there may be particles
escaping the PWN and the particle spectrum in the outer parts of the PWN becomes steeper, which
may be the case for old ($\gapp10^5\ \rm yr$) PWNe.
For young PWNe, TC12 uses a reflecting boundary condition at the outer edge of the PWN.
We note that the TC12 model is for spherically symmetric PWNe and may not be optimal for MSH~15$-$5{\sl2}.
However, the azimuthal variation in the northern nebula is not large (see Fig.~\ref{fig:AzVar}), and thus
the model may provide a reasonable description of the source in that region.

In this model, the angular size of the `flat' region where the radial
profile of the spectral index is flat
can be used to estimate the diffusion coefficient (Equation 14 of TC12).
This is calculated using the following equations:
$$\theta_{\rm flat}\approx \theta\left(\frac{6^{1/2}}{2}\left[\frac{\nu_R}{\nu}\right]^{1/4} -1\right)$$
and
$$\nu_{R}=1\times 10^{17}\left(\frac{D}{10^{27}\ \rm cm^2\ s^{-1}}\right)^2
\left(\frac{1\ \rm pc}{R_{\rm PWN}}\right)^4\left(\frac{100 \mu\rm G}{B}\right)^3\ \rm Hz,$$
where $\theta_{\rm flat}$ is the angular size of the region
that has a flat photon index profile, $\theta$ is the angular size of the PWN,
$\nu$ is the photon frequency, and $D$ is the diffusion coefficient.

We find that photon index profile steepens more slowly beyond $R=71''$ and $R=110''$
in the northern and the jet regions, respectively. Note that the radial profile of the spectral
index in the northern nebula shows a flat region between $R=70''$--$200''$ although the profile
seems not to show any flat region in the southern nebula.
We use the value for the northern nebula for the size of the flat region of TC12.
Assuming the size of the source is $R_{\rm PWN}\sim300''$ and
using the above formulae with $\nu=2.4\times10^{17}$~Hz (1~keV),
we estimated the diffusion coefficient to be
4--13$\times 10^{27}\ \rm cm^2\ s^{-1}$ for $B$=8--17 $\mu$G, which is slightly
larger than that estimated for 3C~58 by numerical modeling ($2.9\times10^{27}\ \rm cm^2\ s^{-1}$, TC12).

Using the diffusion coefficient we estimated above,
we calculate the critical particle energy $E_{\rm R}$ for which the diffusion distance
is equal to the size of the PWN (see Table~3 of TC12) and where the electron distribution
has a break \citep[][]{g72}, using formulae given by TC12:
$R=(4D/QE_R)^\frac{1}{2}$ and $Q=1.58\times 10^{-3} B^2\ \rm erg\ s^{-1}$.
For MSH~15$-$5{\sl2}, we find $E_R$ to be 130--190 TeV for $B$=8--17 $\mu$G.
It is interesting to note that this is similar to the maximum electron energies inferred
from broadband SED modeling \citep[130 or 250 TeV;][]{nky+08},
and that inferred from the \AR{possible} spectral break at 6~keV we measured in Section~\ref{sec:totspec}.

We note that the spectrum in the jet direction is significantly softer in the innermost J1
region compared to that in the farther J2 region. Such behavior is not expected
in simple advection and/or diffusion models,
since the synchrotron emitting particle spectrum only softens with distance. This simple picture
may not be appropriate in the regions where the particle flow may be more complicated due to
magnetic hoop stress as suggested for this source by \citet{yks+09}.
The authors found a ring-like structure with $R\sim$10$''$ using {\em Chandra} data
and interpreted the structure as the
termination shock for this PWN. Based on the morphology, the authors further suggested that
the shock accelerated particles are diverted and squeezed towards the poloidal direction
right below the ring due to magnetic hoop stress \citep[e.g.,][]{l02}.
We also find that the jet structure becomes narrower to $R\sim35''$ and then broader. Furthermore,
the spectral hardness turn-over, non-monotonic variation of the spectral index (see Fig.~\ref{fig:RadVar}),
happens near the location where the jet is narrowest,
which might be occurring because of the compression of the magnetic fields and particles.

\medskip
\subsection{The 2-D Spectral Maps}
\label{sec:2Dnh}
We presented 2-D maps of the spectral parameters. The maps visualize the properties of the source
very well, and can be compared with 3-D PWN models.

We showed that the 2-D map of $N_{\rm H}$ has a shell-like
structure. The density is low near the central pulsar, increasing out to $R\sim3'$
(see also Figs.~\ref{fig:AzVar} and \ref{fig:RadVar}).
We note that the fit value of $N_{\rm H}$ could in principle
be degenerate with other spectral parameters.
However, we do not find clear evidence of
correlation between $N_{\rm H}$ and photon index or flux from our analysis (see Section~\ref{sec:2dmap}),
and using a constant $N_{\rm H}$ degrades the fit significantly.

A higher column density is observed in the south and the east directions (see Fig.~\ref{fig:2Dmap}).
If the material responsible for $N_{\rm H}$ was produced by the supernova, one would expect
the pulsar to have a kick in the opposite direction of the material, towards the north-east direction,
which is consistent with the direction of the kick velocity for PSR~B1509$-$58
estimated by \citet[][]{r04} based on 2800 days of timing. However, \citet[][]{lk11} found
no evidence of proper motion using 28 years of timing data.
Nevertheless, we do not see any enhanced emission in the shell-like structure,
which makes the supernova ejecta scenario less plausible.

Alternatively, the structure may be an interstellar bubble produced by the stellar
wind of the supernova progenitor \citep[e.g.,][]{cmw75}.
In the wind model, the size of the bubble
is given by a simple formula:
$$R_s(t)=28\left(\frac{\dot M_6 V_{2000}^2}{n_0}\right)^{1/5} t_6^{3/5}\ \rm pc,$$
where $\dot M_6$ is the mass loss rate of the progenitor in units of $10^{-6} M_\odot \rm yr^{-1}$,
$V_{2000}$ is the speed of the wind in units of $2000\ \rm km\ s^{-1}$, $n_0$ is the number
density ($\rm cm^{-3}$) of the interstellar medium, and $t_6$ is the time in units of $10^6\ \rm yr$.
The radius of the ring structure we observe is $\sim$5 pc, much smaller than the calculated value
for a typical O6 star. However, the value can vary significantly for different input parameters, and
due to spatial non-uniformity of the interstellar medium or radiative loss \citep[e.g.,][]{wmc+77}.

The asymmetric shell structure may be explained by outbursts of massive stars.
The massive star progenitors of core-collapse supernovae undergo a variety of instabilities
that drive episodic mass loss: e.g., pulsations driven by bumps in
the continuum opacity \citep{ln02, fry06, pca+13}
and explosive shell burning \citep{qs12, amv14}.
These outbursts occur up until the collapse of the star and are believed to have large asymmetries.
An outburst a few thousand years prior to collapse could explain the features
we observe at 5 pc in the remnant.

If, as suggested above, the shell-like structure at R$\sim$3$'$
was formed by the stellar wind,
the structure would have to avoid being swept up or destroyed by the SN ejecta.
If the supernova ejecta did not fill a full spherical shell,
a part of the wind-produced shell can be left over. In this case, density of the shell
is expected to be higher in the direction where the supernova ejecta were less dense.
We see such a trend when comparing our
$N_{\rm H}$ map with the radio image of the SNR \citep[Figs.~2 and 3 of][]{gbm+99}; there are
more ejecta in the northern region than in the southern region.

We have estimated the mass of hydrogen contained in the observed shell-like structure. Using
the measured radial profile of $N_{\rm H}$ shown in Figure~\ref{fig:RadVar}d, the excess mass
compared to the central region  is $\sim$460$M_\odot$, large compared to the
$\sim$12$M_\odot$ one would estimate for a sphere with $R=5\ \rm pc$ for typical interstellar
density of $1\ \rm cm^{-3}$.
Furthermore, the large amount of material in the structure should
produce H{\scriptsize I} emission,
which we do not see in the 20 cm map \citep[Fig.~4 of][]{gbm+99}.
This may be because the radio continuum emission was not subtracted in the radio map and/or
because the X-ray measurement is sensitive only to foreground material
while the radio observations are sensitive to both foreground and background structures.

We note that we cannot unambiguously rule out the possibility
of a constant $N_{\rm H}$ over the field; the observed $N_{\rm H}$ being
an artifact of a more complex underlying continuum.
Thus, it is very difficult to clearly interpret the structure
using X-ray observations only.
Nevertheless, if the shell-like structure in the $N_{\rm H}$ map is intrinsic to the source,
it may support the idea of the existence of an underdense region around
the supernova progenitor, which was suggested to explain the discrepancy between the pulsar's
characteristic age of $\sim$1700 yr \citep[][]{kms+94} and the SNR age
of $>$10000 yr based on the RCW~89 association \citep[][]{shm+83}.
This requires further confirmation by observations in other bands.

\medskip
\section{Summary}
\label{sec:concl}
We have presented energy-resolved images of the PWN MSH~15$-$5{\sl2}
in the hard X-ray band ($E>8\ \rm keV$) for the first time.
The images in different X-ray bands shrink with energy as a result of
the synchrotron burn-off effect.
On small scales ($R\lapp$50$''$), we show that the size shrinkage with energy
can be explained with a particle advection model.
Using this model, we discuss properties of the wind outflow in the jet direction.
We find \AR{that the combined {\em NuSTAR}/{\em Chandra} spectrum of the entire PWN
requires a break at 6~keV, which may be due to cross-calibration effects.}
\AR{However, if the spectral break is intrinsic to the source,
it} implies a break in the shock accelerated electron distribution. We measured the spectral
index profiles on large scales ($R\sim$5$'$) in the northern and jet directions.
The spectrum softens with radius in both directions, an effect we interpret with
a combined diffusion/advection model; further numerical simulations
with the model are required for more accurate interpretation.
We find an interesting sinusoidal variation of the spectral hardness
in the azimuthal direction which may have implications for the particle diffusion
in the PWN. Such a variation has not been seen in other PWNe, though it has been
predicted in pulsar current flow models \citep[][]{b02}.
We find a spectral hardness turn-over in the jet direction at a distance of $\sim$35$''$
from the pulsar. Finally, we presented 2-D maps of spectral parameters
of the source, and find that the $N_{\rm H}$ map shows
an interesting shell-like structure which implies high particle density. However, this
feature could result from a complex underlying continuum, and so requires further confirmation.\\

\medskip

This work was supported under NASA Contract No. NNG08FD60C, and  made use of data from the {\it NuSTAR} mission,
a project led by  the California Institute of Technology, managed by the Jet Propulsion  Laboratory,
and funded by the National Aeronautics and Space  Administration. We thank the {\it NuSTAR} Operations,
Software and  Calibration teams for support with the execution and analysis of  these observations.
This research has made use of the {\it NuSTAR}  Data Analysis Software (NuSTARDAS) jointly developed by
the ASI  Science Data Center (ASDC, Italy) and the California Institute of  Technology (USA).
V.M.K. acknowledges support
from an NSERC Discovery Grant and Accelerator Supplement,
the FQRNT Centre de Recherche Astrophysique du Qu\'ebec,
an R. Howard Webster Foundation Fellowship from the Canadian Institute for Advanced
Research (CIFAR), the Canada Research Chairs Program and the Lorne Trottier Chair
in Astrophysics and Cosmology.

\end{document}